\documentclass[review]{elsarticle}
\usepackage[margin=3cm]{geometry}

\usepackage[hidelinks,breaklinks=true]{hyperref}

\usepackage{caption}
\usepackage{subcaption}
\usepackage{acro}
\acsetup{first-style=short}
\usepackage{dblfloatfix}
\usepackage{amsmath}
\usepackage[table]{xcolor}
\usepackage{graphicx}
\graphicspath{{./figures/}}
\usepackage{booktabs}
\usepackage{siunitx}
\usepackage[utf8]{inputenc}
\usepackage{csquotes}
\usepackage{breakcites}
\journal{International Journal of Multiphase Flow}

\biboptions{authoryear}

\usepackage{pgfplots}
\newcommand{\Cnoz}{C_{\mathrm{noz}}}
\newcommand{\Lnoz}{\mathsf{L_{noz}}}
\newcommand{\rnoz}{\mathsf{r}}
\newcommand{\Dsac}{\mathsf{D_{sac}}}
\newcommand{\Dnoz}{\mathsf{D_{noz}}}
\newcommand{\Do}{\mathsf{D_o}}
\newcommand{\Uo}{\mathsf{U_o}}
\newcommand{\tauo}{\mathsf{\tau_o}}
\newcommand{\Tlo}{\mathsf{T_{\mathrm{l,o}}}}
\newcommand{\Doo}{\mathsf{D^2_o}}
\newcommand{\Red}{\mathsf{Re}_{\mathrm{{d}}}}
\newcommand{\BMd}{\mathsf{B_{M,\mathrm{{d}}}}}
\newcommand{\Shd}{\mathsf{Sh_{\mathrm{{d}}}}}
\newcommand{\Scg}{\mathsf{Sc_{\mathrm{{g}}}}}
\newcommand{\Prg}{\mathsf{Pr_{\mathrm{{g}}}}}
\newcommand{\Nud}{\mathsf{Nu_{\mathrm{{d}}}}}

\newcommand{\mean}[1]{\langle #1 \rangle}
\newcommand{\bigmean}[1]{\left\langle #1 \right\rangle}

\newcommand{\tsup}[1]{^{\mathrm{#1}}}

\newcommand{\wenta}{\dot{\omega}_{\mathrm{ent,\air}}}
\newcommand{\wentg}{\dot{\omega}_{\mathrm{ent,\gas}}}
\newcommand{\wdrag}{\dot{\omega}_{\mathrm{drag}}}
\newcommand{\wvap}{\dot{\omega}_{\mathrm{vap}}}
\newcommand{\wbre}{\dot{\omega}_{\mathrm{bre}}}
\newcommand{\wheat}{\dot{\omega}_{\mathrm{heat}}}

\newcommand{\gas}{\mathrm{g}}
\newcommand{\air}{\mathrm{a}}
\newcommand{\liq}{\mathrm{l}}
\newcommand{\vap}{\mathrm{v}}
\newcommand{\alphal}{\alpha_{\liq}}

\newcommand{\Dphase}{\mathrm{D_{phase}}}
\newcommand{\Dg}{\mathrm{D_g}}
\newcommand{\Dl}{\mathrm{D_l}}
\newcommand{\rhohat}{\bar{\rho}}
\newcommand{\rhol}{\rho_{\liq}}
\newcommand{\rhog}{\rho_{\gas}}
\newcommand{\rhoa}{\rho_{\air}}
\newcommand{\rhov}{\rho_{\vap}}
\newcommand{\Yhatl}{\hat{Y}_{\liq}}
\newcommand{\Yhatg}{\hat{Y}_{\gas}}
\newcommand{\Yhata}{\hat{Y}_{\air}}
\newcommand{\Yhatv}{\hat{Y}_{\vap}}

\newcommand{\Ytvs}{{Y}_{\mathrm{v,s}}}
\newcommand{\Yiv}{Y_{\mathrm{\infty,v}}}
\newcommand{\chieqvs}{\chi_{\mathrm{eq,v,s}}}
\newcommand{\chineqvs}{\chi_{\mathrm{neq,v,s}}}
\newcommand{\LK}{L_\mathrm{Kn}}
\newcommand{\Wv}{W_{\vap}}
\newcommand{\pv}{p_{\vap}}
\newcommand{\Pg}{P_{\gas}}

\newcommand{\uhatphase}{\hat{u}_{\mathrm{phase}}}
\newcommand{\uhatl}{\hat{u}_{\liq}}
\newcommand{\uhatg}{\hat{u}_{\gas}}

\newcommand{\urel}{\hat{u}_{\mathrm{rel}}}
\newcommand{\mug}{\mu_{\gas}}
\newcommand{\mua}{\mu_{\air}}
\newcommand{\mul}{\mu_{\liq}}
\newcommand{\sigmal}{\sigma_{\liq}}
\newcommand{\ucuberel}{\hat{u}^3_{\mathrm{rel}}}
\newcommand{\dhat}{\hat{d}}

\newcommand{\dst}{d_{\mathrm{st}}}
\newcommand{\taub}{\tau_{\mathrm{b}}}
\newcommand{\dstRD}{d_{\mathrm{st,RD}}}
\newcommand{\taubRD}{\tau_{\mathrm{b,RD}}}
\newcommand{\dstKH}{d_{\mathrm{st,KH}}}
\newcommand{\taubKH}{\tau_{\mathrm{b,KH}}}
\newcommand{\dstRT}{d_{\mathrm{st,RT}}}
\newcommand{\taubRT}{\tau_{\mathrm{b,RT}}}
\newcommand{\Thatd}{\hat{T}_{\mathrm{d}}}
\newcommand{\Td}{{T}_{\mathrm{d}}}
\newcommand{\Tref}{T_{\mathrm{ref}}}
\newcommand{\Tsat}{T_{\mathrm{sat}}}
\newcommand{\Tamb}{T_{\mathrm{amb}}}
\newcommand{\Thatg}{\hat{T}_{\gas}}
\newcommand{\Thata}{\hat{T}_{\air}}
\newcommand{\Thatv}{\hat{T}_{\vap}}
\newcommand{\Cpg}{C_{\mathrm{p,\gas}}}
\newcommand{\Cpa}{C_{\mathrm{p,\air}}}
\newcommand{\Cpv}{C_{\mathrm{p,\vap}}}
\newcommand{\Kbre}{K_{\mathrm{bre}}}
\newcommand{\Cb}{C_{\mathrm{b}}}
\newcommand{\Kvap}{K_{\mathrm{vap}}}
\newcommand{\Gvg}{\mathit{\Gamma}_{\mathrm{v,g}}}
\newcommand{\Kheat}{K_{\mathrm{heat}}}
\newcommand{\Qd}{Q_{\mathrm{d}}}
\newcommand{\Cl}{C_{\liq}}
\newcommand{\lambdag}{\lambda_{\gas}}
\newcommand{\Cd}{C_{\mathrm{D}}}
\newcommand{\Cdrag}{C_{\mathrm{drag}}}
\newcommand{\Cvap}{C_{\mathrm{vap}}}
\newcommand{\Z}{\mathsf{Z}}

\newcommand{\rhostar}{\rho^{*}}
\newcommand{\Ystarl}{{Y}^{*}_{\liq}}
\newcommand{\Ystara}{{Y}^{*}_{\air}}
\newcommand{\Ystarv}{{Y}^{*}_{\vap}}
\newcommand{\Ystarg}{{Y}^{*}_{\gas}}

\newcommand{\ustarl}{{u}^{*}_{\liq}}
\newcommand{\ustarg}{{u}^{*}_{\gas}}

\newcommand{\dstar}{{d}^*}
\newcommand{\dstarsq}{{d}^{*2}}
\newcommand{\Tstard}{{T}^*_{\mathrm{d}}}
\newcommand{\Tstarg}{{T}^*_{\gas}}
\newcommand{\zstar}{{z}^*}
\newcommand{\tstar}{{t}^*}
\newcommand{\bstar}{{b}^*}

\newcommand{\ndo}{\textit{n}-dodecane}
\newcommand{\dnbe}{di-\textit{n}-butylether}
\newcommand{\oct}{1-octanol}

\newcommand{\Pinj}{P_{\mathrm{inj}}}
\newcommand{\Tinj}{T_{\mathrm{inj}}}
\newcommand{\Pamb}{P_{\mathrm{amb}}}

\newcommand{\lp}{\left(}
\newcommand{\rp}{\right)}

\newcommand{\pdo}[2][]{\dfrac{\partial}{\partial {#1}} \lp {#2} \rp}

\newcommand{\myeq}[1]{Eq.~\eqref{#1}}
\newcommand{\myfig}[1]{Figure~\ref{#1}}
\newcommand{\mytab}[1]{Table~\ref{#1}}
\newcommand{\mysec}[1]{Section~\ref{#1}}
\newcommand{\mysubsec}[1]{Subsection~\ref{#1}}

\definecolor{SAEyellow}{rgb}{1,0.698039215686274,0.00392156862745098}
\definecolor{SAEorange}{rgb}{0.917647058823529,0.443137254901961,0.145098039215686}
\definecolor{SAEred}{rgb}{0.862745098039216,0.16078431372549,0.117647058823529}
\definecolor{SAEblue}{rgb}{0.00392156862745098,0.627450980392157,0.913725490196078}
\definecolor{SAEdblue}{rgb}{0,0.317647058823529,0.584313725490196}
\definecolor{SAEgreen}{rgb}{0.180392156862745,0.694117647058824,0.207843137254902}
\definecolor{SAEdgreen}{rgb}{0,0.466666666666667,0.23921568627451}
\definecolor{SAElgray}{rgb}{0.792156862745098,0.792156862745098,0.784313725490196}
\definecolor{SAEmgray}{rgb}{0.603921568627451,0.607843137254902,0.615686274509804}
\definecolor{SAEdgray}{rgb}{0.380392156862745,0.384313725490196,0.396078431372549}

\usepackage{array,multirow,tabularx,ragged2e}
\newcolumntype{L}[1]{>{\raggedright\let\newline\\\arraybackslash\hspace{0pt}}m{#1}}
\newcolumntype{C}[1]{>{\centering\let\newline\\\arraybackslash\hspace{0pt}}m{#1}}
\newcolumntype{R}[1]{>{\raggedleft\let\newline\\\arraybackslash\hspace{0pt}}m{#1}}

\DeclareAcronym{GHG}{
  short = GHG ,
  long  = Greenhouse Gas ,
  class = abbrev
}

\DeclareAcronym{INGAS}{
  short = INGAS ,
  long  = Integrated Gas Powertrain ,
  class = abbrev
}

\DeclareAcronym{ICE}{
  short = ICE ,
  long  = Internal Combustion Engine ,
  class = abbrev
}

\DeclareAcronym{SI}{
  short = SI ,
  long  = Spark Ignition ,
  class = abbrev
}

\DeclareAcronym{CI}{
  short = CI ,
  long  = Compression Ignition ,
  class = abbrev
}

\DeclareAcronym{CNG}{
  short = CNG ,
  long  = Compressed Natural Gas ,
  class = abbrev
}
\DeclareAcronym{NG}{
  short = NG ,
  long  = Natural Gas ,
  class = abbrev
}
\DeclareAcronym{PFI}{
  short = PFI ,
  long  = Port Fuel Injection ,
  class = abbrev
}
\DeclareAcronym{PLIF}{
  short = PLIF ,
  long  = Planar Laser-Induced Fluorescence ,
  class = abbrev
}
\DeclareAcronym{DI}{
  short = DI ,
  long  = Direct Injection ,
  class = abbrev
}

\DeclareAcronym{CoV}{
  short = CoV ,
  long  = Coefficient of Variation ,
  class = abbrev
}

\DeclareAcronym{HC}{
  short = HC ,
  long  = Hydrocarbon ,
  class = abbrev
}

\DeclareAcronym{EUCAR}{
  short = EUCAR ,
  long  = European Council for Automotive Research and Development ,
  class = abbrev
}
\DeclareAcronym{LES}{
  short = LES ,
  long  = Large-Eddy Simulation ,
  class = abbrev
}
\DeclareAcronym{DNS}{
  short = DNS ,
  long  = Direct Numerical Simulation ,
  class = abbrev
}
\DeclareAcronym{MBC}{
  short = MBC ,
  long  = Mapped Boundary Condition ,
  class = abbrev
}
\DeclareAcronym{3-D}{
  short = 3-D ,
  long  = Three-dimensional ,
  class = abbrev
}
\DeclareAcronym{3D}{
  short = 3D ,
  long  = Three-dimensional ,
  class = abbrev
}
\DeclareAcronym{2-D}{
  short = 2-D ,
  long  = Two-dimensional,
  class = abbrev
}
\DeclareAcronym{2D}{
  short = 2D ,
  long  = Two-dimensional,
  class = abbrev
}
\DeclareAcronym{1-D}{
  short = 1-D ,
  long  = One-dimensional,
  class = abbrev
}
\DeclareAcronym{1D}{
  short = 1D ,
  long  = One-dimensional,
  class = abbrev
}
\DeclareAcronym{PIV}{
  short = PIV ,
  long  = Particle-Image Velocimetry,
  class = abbrev
}
\DeclareAcronym{TVD}{
  short = TVD ,
  long  = Total Variation Diminishing,
  class = abbrev
}
\DeclareAcronym{MUSCL}{
  short = MUSCL,
  long  = Monotonic Upwind Scheme for Conservation Laws,
  class = abbrev
}
\DeclareAcronym{CHRIS}{
  short = CHRIS,
  long  = Compressible High-speed Reactive Solver,
  class = abbrev
}
\DeclareAcronym{NPR}{
  short = NPR,
  long  = Nozzle Pressure Ratio,
  class = abbrev
}
\DeclareAcronym{CFD}{
  short = CFD,
  long  = Computational Fluid Dynamics,
  class = abbrev
}
\DeclareAcronym{PISO}{
  short = PISO,
  long  = Pressure Implicit with Splitting Operators,
  class = abbrev
}
\DeclareAcronym{RANS}{
  short = RANS,
  long  = Reynolds-Averaged Navier-Stokes,
  class = abbrev
}
\DeclareAcronym{URANS}{
  short = URANS,
  long  = Unsteady Reynolds-Averaged Navier-Stokes,
  class = abbrev
}
\DeclareAcronym{RNG}{
  short = RNG,
  long  = Re-Normalized Grouping,
  class = abbrev
}
\DeclareAcronym{UDF}{
  short = UDF,
  long  = User Defined Function,
  class = abbrev
}
\DeclareAcronym{DCS}{
  short = DCS,
  long  = Dynamically Coupled Source,
  class = abbrev
}
\DeclareAcronym{AMR}{
  short = AMR,
  long  = Adaptive Mesh Refinement,
  class = abbrev
}
\DeclareAcronym{CFL}{
  short = CFL,
  long  = Courant-Friedrichs-Lewy,
  class = abbrev
}
\DeclareAcronym{APL}{
  short = APL,
  long  = Axial Penetration Length,
  class = abbrev
}
\DeclareAcronym{MW}{
  short = MW,
  long  = Maximum Width,
  class = abbrev
}
\DeclareAcronym{AJ}{
  short = AJ,
  long  = Area of Jet,
  class = abbrev
}
\DeclareAcronym{VJ}{
  short = VJ,
  long  = Volume of Jet,
  class = abbrev
}
\DeclareAcronym{TDC}{
  short = TDC,
  long  = Top Dead Center,
  class = abbrev
}
\DeclareAcronym{ATDC}{
  short = ATDC,
  long  = After Top Dead Center,
  class = abbrev
}
\DeclareAcronym{SOI}{
  short = SOI,
  long  = Start of Injection,
  class = abbrev
}
\DeclareAcronym{EOI}{
  short = EOI,
  long  = End of Injection,
  class = abbrev
}
\DeclareAcronym{CA}{
  short = CA,
  long  = Crank Angle,
  class = abbrev
}
\DeclareAcronym{RPM}{
  short = RPM,
  long  = Rotations per Minute,
  class = abbrev
}
\DeclareAcronym{TKE}{
  short = TKE,
  long  = Turbulent Kinetic Energy,
  class = abbrev
}
\DeclareAcronym{MC}{
  short = MC,
  long  = Monotonic Center,
  class = abbrev
}
\DeclareAcronym{PDE}{
  short = PDE,
  long  = Partial Differential Equation,
  class = abbrev
}
\DeclareAcronym{CDV}{
  short = CDV,
  long  = Converging-Diverging Verification,
  class = abbrev
}
\DeclareAcronym{CAS}{
  short = CAS ,
  long  = Cross-sectionally Averaged Spray ,
  class = abbrev
}
\DeclareAcronym{RD}{
  short = RD ,
  long  = Reitz-Diwaker ,
  class = abbrev
}
\DeclareAcronym{KHRT}{
  short = KH-RT ,
  long  = Kelvin-Helmholtz Rayleigh-Taylor ,
  class = abbrev
}
\DeclareAcronym{KH}{
  short = KH ,
  long  = Kelvin-Helmholtz ,
  class = abbrev
}
\DeclareAcronym{RT}{
  short = RT ,
  long  = Rayleigh-Taylor ,
  class = abbrev
}
\DeclareAcronym{WENO}{
  short = WENO ,
  long  = Weighted Essentially Non-oscillatory ,
  class = abbrev
}
\DeclareAcronym{ND}{
  short = ND ,
  long  = Non-dimensional ,
  class = abbrev
}
\DeclareAcronym{BC}{
  short = BC ,
  long  = Boundary Condition ,
  class = abbrev
}
\DeclareAcronym{IC}{
  short = IC ,
  long  = Initial Condition ,
  class = abbrev
}
\DeclareAcronym{ECN}{
  short = ECN ,
  long  = Engine Combustion Network ,
  class = abbrev
}
\DeclareAcronym{FSC}{
  short = FSC ,
  long  = Fuel Science Center ,
  class = abbrev
}
\DeclareAcronym{PDF}{
  short = PDF,
  long  = Probability Density Function ,
  class = abbrev
}
\DeclareAcronym{DBI}{
  short = DBI,
  long  = Diffuse Background Illumination ,
  class = abbrev
}

\DeclareAcronym{CO2}{
  short = CO$_2$,
  long  = Carbon dioxide ,
  sort  = CO2 ,
  class = nomencl
}
\DeclareAcronym{CH4}{
  short = CH$_4$,
  long  = Methane ,
  sort  = CH4 ,
  class = nomencl
}
\DeclareAcronym{N2}{
  short = N$_2$,
  long  = Nitrogen ,
  sort  = N2 ,
  class = nomencl
}
\DeclareAcronym{O2}{
  short = O$_2$,
  long  = Oxygen ,
  sort  = O2 ,
  class = nomencl
}
\DeclareAcronym{Spsi}{
  short = $S_{\Psi}$ ,
  long  = Source term for flux $\Psi$ ,
  sort  = Spsi ,
  class = nomencl
}
\DeclareAcronym{Psi}{
  short = $\Psi$ ,
  long  = Flux ,
  sort  = Psi ,
  class = nomencl
}
\DeclareAcronym{mdot}{
  short = $\dot{m}$ ,
  long  = Mass flow rate ,
  sort  = mdot ,
  class = nomencl
}
\DeclareAcronym{z}{
  short = $z$ ,
  long  = Axial coordinate ,
  sort  = z ,
  class = nomencl
}
\DeclareAcronym{r}{
  short = $r$ ,
  long  = Radial coordinate ,
  sort  = r ,
  class = nomencl
}
\DeclareAcronym{varphi}{
  short = $\varphi$ ,
  long  = Azimuthal coordinate ,
  sort  = varphi ,
  class = nomencl
}
\DeclareAcronym{x}{
  short = x ,
  long  = Co-ordinate along the center line ,
  sort  = x ,
  class = nomencl
}
\DeclareAcronym{t}{
  short = $t$ ,
  long  = time coordinate ,
  sort  = t ,
  class = nomencl
}
\DeclareAcronym{A}{
  short = $A$ ,
  long  = Area of cross-section ,
  sort  = A ,
  class = nomencl
}
\DeclareAcronym{rho}{
  short = $\rho$ ,
  long  = Density ,
  sort  = rho ,
  class = nomencl
}
\DeclareAcronym{b}{
  short = $b$ ,
  long  = Maximum radial distance for integration ,
  sort  = b ,
  class = nomencl
}
\DeclareAcronym{u}{
  short = $u$ ,
  long  = Velocity ,
  sort  = u ,
  class = nomencl
}
\DeclareAcronym{p}{
  short = $p$ ,
  long  = Pressure ,
  sort  = p ,
  class = nomencl
}
\DeclareAcronym{E}{
  short = $E$ ,
  long  = Total energy per unit volume ,
  sort  = u ,
  class = nomencl
}
\DeclareAcronym{gamma}{
  short = $\gamma$ ,
  long  = Ratio of specific heats ,
  sort  = gamma ,
  class = nomencl
}
\DeclareAcronym{Ma}{
  short = Ma ,
  long  = Mach number ,
  sort  = Ma ,
  class = nomencl
}
\DeclareAcronym{N}{
  short = N,
  long  = Engine Speed in RPM,
  class = nomencl
}
\DeclareAcronym{omegacrank}{
  short = $\omega_{crank}$,
  long  = Angular speed of the crank,
  class = nomencl
}
\DeclareAcronym{Hz}{
  short = Hz ,
  long  = Hertz ,
  sort  = Hz ,
  class = nomencl
}
\DeclareAcronym{phi}{
  short = $\phi$ ,
  long  = Quantity of interest ,
  sort  = phi ,
  class = nomencl
}
\DeclareAcronym{psi}{
  short = $\psi$ ,
  long  = Spectral radius ,
  sort  = psi ,
  class = nomencl
}
\DeclareAcronym{Uvec}{
  short = {\bf U} ,
  long  = State vector ,
  sort  = Uvec ,
  class = nomencl
}
\DeclareAcronym{Fluxvec}{
  short = {\bf F}({\bf U)} ,
  long  = Flux vector ,
  sort  = Fluxvec ,
  class = nomencl
}
\DeclareAcronym{Svec}{
  short = {\bf S} ,
  long  = Source vector ,
  sort  = Svec ,
  class = nomencl
}
\DeclareAcronym{awav}{
  short = $a$ ,
  long  = Wave propagation speed,
  sort  = awav ,
  class = nomencl
}
\DeclareAcronym{length}{
  short = $l$ ,
  long  = Length,
  sort  = length ,
  class = nomencl
}
\DeclareAcronym{slope}{
  short = $m$ ,
  long  = Slope of line,
  sort  = slope ,
  class = nomencl
}
\DeclareAcronym{ewg}{
  short = $s$ ,
  long  = Effective width of the gap,
  sort  = ewg ,
  class = nomencl
}
\DeclareAcronym{NL}{
  short = $h$ ,
  long  = Needle lift,
  sort  = NL ,
  class = nomencl
}

\DeclareAcronym{l}{
  short = l,
  long  = Liquid phase ,
  sort  = l ,
  class = subscr
}
\DeclareAcronym{g}{
  short = g,
  long  = Gaseous phase ,
  sort  = g ,
  class = subscr
}
\DeclareAcronym{a}{
  short = a,
  long  = Air ,
  sort  = a ,
  class = subscr
}
\DeclareAcronym{v}{
  short = v,
  long  = Vapor ,
  sort  = v ,
  class = subscr
}
\DeclareAcronym{f}{
  short = f,
  long  = Fuel ,
  sort  = f ,
  class = subscr
}
\DeclareAcronym{d}{
  short = d,
  long  = Droplet property ,
  sort  = d ,
  class = subscr
}
\DeclareAcronym{s}{
  short = s,
  long  = Droplet surface property ,
  sort  = s ,
  class = subscr
}
\DeclareAcronym{ent}{
  short = ent,
  long  = Entrainment ,
  sort  = ent ,
  class = subscr
}
\DeclareAcronym{eq}{
  short = eq,
  long  = Equilibrium ,
  sort  = eq ,
  class = subscr
}
\DeclareAcronym{neq}{
  short = neq,
  long  = Non-equilibrium ,
  sort  = neq ,
  class = subscr
}

\begin{document}
\begin{frontmatter}

\title{{A Reduced-order Model for Multiphase Simulation of Transient Inert Sprays}}


\author[mymainaddress]{A. Y. Deshmukh\corref{mycorrespondingauthor}}
\cortext[mycorrespondingauthor]{Corresponding author}
\ead{a.deshmukh@itv.rwth-aachen.de}
\author[mymainaddress]{T. Grenga}
\author[mymainaddress]{M. Davidovic}
\author[mysecondaryaddress]{L. Schumacher}
\author[mysecondaryaddress]{J. Palmer}
\author[mysecondaryaddress]{M. A. Reddemann}
\author[mysecondaryaddress]{R. Kneer}
\author[mymainaddress]{H. Pitsch}

\address[mymainaddress]{Institute for Combustion Technology, RWTH Aachen University, Aachen, Germany}
\address[mysecondaryaddress]{Institute of Heat and Mass Transfer, RWTH Aachen University, Aachen, Germany}

\begin{abstract}
In global efforts to reduce harmful greenhouse gas emissions from the transportation sector, novel bio-hybrid liquid fuels from renewable energy and carbon sources can be a major form of energy for future propulsion systems due to their high energy density. A fundamental understanding of the spray and mixing performance of the new fuel candidates in combustion systems is necessary to design and develop the fuels for advanced combustion concepts. In the fuel design process, a large number of candidates is required to be screened to arrive at potential fuels for further detailed investigations. For such a screening process, three-dimensional (3D) simulation models are computationally too expensive and hence unfeasible. Therefore, in this paper, we present a fast, reduced-order model for inert sprays. The model is based on the cross-sectionally averaged spray (CAS) model derived by~\cite{Wan1997} from 3D multiphase equations. The original model was first tested against a wide range of conditions and different fuels. The discrepancies between the CAS model and experimental data are addressed by integrating state-of-the-art breakup and evaporation models. In addition, a transport equation for vapor mass fraction is proposed, which is important for evaporation modeling. Furthermore, the model is extended to consider polydispersed droplets by modeling the droplet size distribution by commonly used presumed probability density functions, such as Rosin-Rammler, lognormal, and gamma distributions. The improved CAS model is capable of predicting trends in the macroscopic spray characteristics for a wide range of conditions and fuels. The computational cost of the CAS model is lower than the 3D simulation methods by up to 6 orders of magnitude depending on the method. This enables the model to be used not only for the rapid screening of novel fuel candidates, but also for other applications, where reduced-order modeling is useful.
\end{abstract}

\begin{keyword}
%\texttt{elsarticle.cls}\sep \LaTeX\sep Elsevier \sep template
%\MSC[2010] 00-01\sep  99-00
Reduced-order model \sep CAS  \sep inert spray \sep droplet size distribution \sep bio-hybrid fuels
\end{keyword}

\end{frontmatter}

%\linenumbers

%\input{texsrc/highlights.tex}

\section{Introduction}
With increasing concerns about the environment due to rising global temperatures, greenhouse gas (\ac{GHG}) emissions from the transportation sector are among the targets of stringent control as recommended by international governing institutions~\citep{EuropeanCommission2018}. According to the~\cite{IPCC2014} report, the transportation sector contributes approximately \SI{23}{\percent} of total global GHG emissions. This share is going to become larger with increasing demand unless mitigation efforts are made to develop low-carbon fuels and energy-efficient propulsion systems. Burning fossil fuels, such as gasoline and diesel, contributes a vast amount to this energy demand~\citep{USEIA2016}.\par
To reduce direct GHG emissions from internal combustion engines (\ac{ICE}s), powertrain electrification is emerging as an alternative. However, the life-cycle assessment studies of electric vehicles show that such vehicles are still not economically viable and cannot reduce overall well-to-wheel emissions if the electricity is generated mainly from fossil sources~\citep{Rupp2018,Rupp2019,Rupp2020,Helmers2020}. Therefore, a mix of propulsion technologies is required to meet  the well-to-wheel emissions reduction~\citep{Senecal2019}. This implies that ICEs will remain important primary sources of power in conventional as well as hybrid propulsion systems. To reduce their environmental impact, it is necessary to develop carbon-neutral fuels and improve combustion engines through new combustion concepts that enhance efficiency and reduce pollutant emissions.\par 
Global research programs, such as Co-Optimization of Fuels and Engines~\citep{COOPTIMA2020} and~\cite{FES2020} among others, are developing low-carbon fuels and highly efficient engine concepts. Similarly, the Cluster of Excellence,``The Fuel Science Center"~\citep{FSC2020} at RWTH Aachen University, aims to integrate renewable electricity with a combined utilization of carbon dioxide (\ac{CO2}) and carbon sources from ligno-cellulosic biomass to produce new high-density liquid energy carriers dubbed as bio-hybrid fuels. Together with the new bio-hybrid fuels, innovative engine concepts are being developed for highly efficient and clean combustion processes.\par
In the fuel design process, fuels need to be tested for their applicability in conventional as well as new combustion concepts. Such testing and validation of all potential fuel candidates is not always possible at the engine test bench due to several reasons. The new fuel candidates may not be available in large quantities initially, and only molecular structure is known from proposed production processes. Also, combustion systems may need to be adapted before extensive testing. Since the atomization of the liquid fuel jet and its subsequent mixture formation is an essential intermediate step towards combustion and emissions, simplified experiments for spray breakup and mixing can be performed under engine-like conditions in a constant volume or constant pressure chamber to reduce the required amount of fuel. However, even such experiments can be impractical for a large number of candidates for the same reasons as mentioned before. In this case, predictive numerical simulations of spray and mixture formation can be useful for assessing the performance of fuel candidates in engine-like conditions. Such simulations, depending on the method and approach, may be computationally expensive, e.g., direct numerical simulation (\ac{DNS}) of an ICE with a full injection system is intractable with current state-of-the-art computing resources available worldwide. These simulations may be feasible with large-eddy simulation (\ac{LES}) or Reynolds-averaged Navier-Stokes (\ac{RANS}) approach. Still, it may not be practical to perform such simulations for each fuel candidate. Reduced-order models can provide trends and initial estimates to evaluate a large number of potential novel fuel candidates. A small number of fuel candidates can be selected with the initial screening for further more detailed investigations.\par
The development of control strategies is an integral part of the combustion system development. Adaptive control strategies can potentially help to run combustion devices with optimal efficiency in real time to meet performance and emission targets. Adapting to the strongly non-linear nature of the chemical and physical processes in combustion requires sophisticated model-based closed-loop control strategies, e.g., in the context of diesel and dual-fuel engines~\citep{Albrecht2007,Bengtsson2007,Hillion2009,Ritter2017,Korkmaz2018,Korkmaz2018a}. In this case, reduced-order models, acting as a digital twin of the combustion system, can provide physics-based predictions to the controller to adjust the input parameters.\par
Thus, simplified yet reliable physics-based models find application in many areas including but not limited to the rapid screening of novel bio-hybrid fuel candidates and physics-based closed-loop control of the energy conversion devices.
\subsection{State-of-the-art}
The pathway from delivery of the fuel into the combustion engine through to the emission of exhaust gases out of the engine can be divided into several stages: fuel jet primary breakup and atomization; evaporation; mixing of fuel with the ambient gas; ignition; combustion; and pollutant formation. In this paper, we focus on jet primary breakup and atomization, evaporation, and mixing processes. The fuel spray can be characterized in terms of different global parameters, such as liquid and vapor penetration length, droplet size distribution, distribution of fuel vapor mass fractions along the spray axis. These quantities depend on thermo-physical properties of the fuel and directly influence further mixing, combustion, and pollutant formation. Several early studies on diesel sprays investigated macroscopic characteristics of sprays experimentally~\citep{Wakuri1959,Wakuri1960,Hiroyasu1980}  and developed correlations for the spray tip penetration, spray angle, breakup time, and droplet size distribution. Further modeling efforts \citep{Dent1971,Hay1972,Naber1996} developed power laws for the spray tip penetration with respect to time, nozzle exit diameter, the pressure difference across the nozzle, density of fuel, and ambient density. \cite{Sazhin2001} derived approximate analytical expressions for the spray tip penetration in initial stages as well as fully developed flow. The authors modeled the initial stage as the single phase flow dominated by the liquid phase and later stages as the two-phase flow. It was found that the diesel spray could be described solely on the basis of a two-phase approximation throughout all stages, as the fuel jet breaks up immediately after leaving the nozzle. Later, simplified models were developed for the macroscopic characterization of diesel sprays~\citep{Desantes2006}. \cite{Pastor2008} developed a one-dimensional Eulerian model for transient inert sprays based on the assumption of locally homogeneous and mixing-controlled processes. \cite{Desantes2009} extended this model to reactive diesel sprays and later to multicomponent fuels~\citep{Pastor2015}. \cite{Musculus2009} extended the steady-state control volume approach of~\cite{Naber1996} to transient non-vaporizing sprays to study entrainment waves at the end of injection. \cite{Xu2016} extended the steady-state homogeneous control volume approach to consider the heterogeneous distribution of velocity and fuel volume fraction over the cross-section, which improved the prediction of liquid length.\par
\subsection{Scope and Objectives}
Most of the previous studies focused on modeling of diesel jets using simplified approaches derived from conservation of mass, momentum, and energy fluxes across the control volumes along the axis of the spray. With the exception of~\cite{Sazhin2001}, other models neglected droplet dynamics assuming homogeneous two-phase mixture. These models were originally derived for one-dimensional (\ac{1D}) systems. \cite{Wan1997} followed a top-down approach for model reduction and derived a 1D cross-sectionally averaged spray (\ac{CAS}) model for two-phase diesel spray from the three-dimensional (\ac{3D}) multiphase model. The model considered most of the important processes, such as droplet drag, breakup, evaporation, and entrainment, and was applied to both non-vaporizing and vaporizing sprays for prediction of macroscopic spray characteristics~\citep{Wan1997a,Wan1999}. In addition, it was also interactively coupled with a 3D computational fluid dynamics (\ac{CFD}) code to model the near-nozzle dense spray region~\citep{Hasse2002,Hasse2003}. The reduced-order modeling with a description of all physical processes relevant to transient inert sprays makes it particularly suited for the application to a wide range of fuels and conditions, including those where near-nozzle dense spray modeling is important and droplet dynamics cannot be neglected.\par
Along these lines, the objectives of this paper are twofold: (1) to evaluate the original CAS model for modern injection systems and novel fuel candidates; and (2) to propose crucial improvements to the model to increase its applicability to a wide range of conditions and fuels. This paper focuses on inert sprays as the first part of a series of two papers on the reduced-order spray model for inert and turbulent reactive sprays.\\\par
The remaining article is arranged as follows. In~\mysec{sec:new_model}, the new improved CAS model is presented.~\mysec{sec:case_desc} describes the measurement cases, experimental methods, and 3D simulation methods and data used for validation. The results of the original and the newly developed CAS model are discussed in~\mysec{sec:results}. Finally, the paper finishes with a summary and conclusions. For completeness and further reference, the original CAS model and some results are included in~\ref{sec:app_org_model}.

\section{Models and Methods}\label{sec:new_model}
Inspired by the analysis of jet flames,~\cite{Wan1997} proposed a radially integrated multiphase formulation for non-vaporizing and vaporizing sprays, termed as CAS model, which has been shown to work well for diesel-like engine conditions~\citep{Wan1997a,Wan1999,Hasse2002,Hasse2003}. Under the boundary layer assumptions, the density-weighted cross-sectional average of the quantity of interest, \ac{phi}, is defined as
%\begin{linenomath*}
\begin{align}
	\rhohat \hat{\phi} b^2 = 2 \int^{\infty}_0 \rho \phi r \mathrm{d}r\,,
	\label{eq:radint}
\end{align}
%\end{linenomath*}
where \ac{rho} is the density, \ac{b}$(z,t)$ the spray half-width, \ac{r} the radial coordinate, \ac{z} the axial coordinate, and \ac{t} the temporal coordinate. The `overline' operator $\bar{\cdot}$ and the `hat' operator $\hat{\cdot}$ denote cross-sectional averaging (with $\phi=1$) and density-weighted cross-sectional averaging, respectively, as defined in \myeq{eq:radint}. The radially integrated multiphase differential operator is defined for each phase as
%\begin{linenomath*}
\begin{align}
        \Dphase \left(\rhohat\hat{\phi}b^2\right) &= \pdo[t]{\rhohat\hat{\phi}b^2}+ \pdo[z]{\rhohat\hat{\phi}\uhatphase b^2}~ \text{with phase = g, l} \,,
        \label{eq:operator}
\end{align}
%\end{linenomath*}
where subscripts `\ac{g}' and `\ac{l}' denote gaseous and liquid phase, respectively. All quantities are assumed to be uncorrelated in the radial direction, which simplifies the evaluation of radial integrals of joint quantities.
\subsection{Governing Equations}
In the original work of~\cite{Wan1997}, multi-dimensional equations for a complete spray~\citep{Hiroyasu1980} were radially integrated, assuming boundary layer approximations, azimuthal symmetry, and top-hat profiles for all flow variables. A detailed description of the original CAS model including the models for droplet breakup and evaporation and the comparison of its performance with respect to LES and URANS simulations is reported in \ref{sec:app_org_model}. The original governing equations for the CAS model (Eq. \eqref{eq:old_gas_cont}-\eqref{eq:old_liq_energy}) did not explicitly compute the vaporized fuel mass fraction, which can have a significant impact on the evaporation process through the change in gas mixture properties. The spray modeling has progressed significantly since the CAS model was first proposed in 1997, especially in the spray breakup and evaporation modeling. In this work, several advances are proposed to improve the model predictions for the current fuel injection systems. Current state-of-the-art models for spray breakup and evaporation models were integrated into the CAS model. An additional equation is solved for the fuel-vapor transport, which is essential for the accurate modeling of evaporation rates. Here, a new approach is developed to consider different droplet sizes instead of just assuming that the spray is monodisperse. The new improved CAS model is described by the following system of hyperbolic partial differential equations (\ac{PDE}s) with source terms: 
%\begin{linenomath*}
\begin{align}
		\Dg( \rhohat \Yhata b^2) & = \wenta b \label{eq:amb_cont} \\
		\Dg( \rhohat \Yhatv b^2) & =\mean{\wvap} b^2 \label{eq:vap_cont} \\
		\Dg(\rhohat\Yhatg \uhatg b^2) & = -\mean{\wdrag} b^2 + \mean{\wvap} \uhatl  b^2 \label{eq:gas_mom} \\
		\Dl(\rhohat\Yhatl b^2) & =  - \mean{\wvap} b^2 \label{eq:liq_cont} \\
		\Dl(\rhohat\Yhatl \uhatl b^2) & = \mean{\wdrag}b^2 - \mean{\wvap} \uhatl  b^2 \label{eq:liq_mom} \\
		\Dl (\rhohat\Yhatl \mean{\hat{d^2}} b^2) & = -\mean{\wbre} b^2 - \frac{5}{3}\mean{\wvap \dhat^2 } b^2 \label{eq:d2_trans}\\
		\Dl (\rhohat\Yhatl \mean{\dhat} b^2) & = -\bigmean{\frac{\wbre}{2\dhat}} b^2 - \frac{4}{3}\mean{\wvap\dhat} b^2 \label{eq:d_trans} \\
		\Dl(\rhohat\Yhatl \Thatd b^2) & =   \mean{\wheat} b^2 - \mean{\wvap  \Thatd} b^2 \label{eq:liq_energy}
		%\label{eq:cas_new}
\end{align}
%\end{linenomath*}
The quantity $\phi$ has been replaced by individual or combinations of the flow variables, such as the mass fractions ($Y_{\air,\vap,\liq}$) and the velocities ($u_{\gas,\liq}$) of the respective phases, the droplet diameter ($d$), and the droplet temperature ($\Td$) to obtain the set of PDEs. The subscript `\ac{d}' marks the droplet variables. Any term, $\zeta(\dhat)$, which depends on the droplet diameter, is integrated over the droplet size distribution, $\mathcal{P}(\dhat)$, as
%\begin{linenomath*}
\begin{align}
	\mean{\zeta} = \int \zeta(\dhat') \mathcal{P}(\dhat') \mathrm{d} \dhat'\,,
	\label{eq:pdfint}
\end{align}
%\end{linenomath*}
where the operator, $\mean{\cdot}$, defines the expectation value of $\zeta(\dhat)$.
The subscripts `\ac{a}'  and `\ac{v}' denote ambient gas and vapor, respectively. Mass conservation requires $\Yhatl+\Yhatv+\Yhata = 1$ and $\Yhatg = \Yhatv+\Yhata$. The two-phase cross-sectionally averaged mixture density is computed as
%\begin{linenomath*}
\begin{align}
		\frac{1}{\rhohat} & = \frac{\Yhatl}{\rhol} + \frac{\Yhatv}{\rhov} + \frac{\Yhata}{\rhoa}\,,
		\label{eq:mix_den}
\end{align} 
%\end{linenomath*}
where $\rho_{\liq,\vap,\air}$ denote densities of liquid, vapor, and ambient gas, respectively. The spray half-width, $b(z,t)$, is computed by local mass conservation by summation of~\myeq{eq:amb_cont},~\myeq{eq:vap_cont}, and~\myeq{eq:liq_cont} and then using the mixture density $\rhohat$. The source terms on the right-hand side, represent different models for physical processes, namely, entrainment ($\wenta$), drag ($\wdrag$), evaporation ($\wvap$), droplet breakup ($\wbre$), and droplet heating ($\wheat$). The physical sub-models used are described briefly in the next subsections.
\subsection{Entrainment Model}\label{sec:ent_model}
Entrainment of the ambient gas into the spray plume leads to its spreading in the radial direction, which develops the spray cone angle. The source term for entrainment is modeled as $\wenta = \rhoa \beta \uhatg$ with spreading coefficient, $\beta = \mathrm{tan}(\theta/2)$, depending on the spray cone angle, $\theta$. Several correlations for the spray cone angle were discussed by~\cite{Wan1997}, eventually using values interpolated from~\cite{Naber1996}. In this study, the correlation by~\cite{Hiroyasu1980} is used, which considers the effects of nozzle internal geometry on the spray cone angle, neglecting cavitation effects and has been shown to closely match with the experimental data. The spray cone angle ($\theta$, in degree) is given as
%\begin{linenomath*}
\begin{align}
		\theta = \Cnoz \left(\frac{\rhoa}{\rhol}\right)^{0.26}\,,
		\label{eq:theta}
\end{align} 
%\end{linenomath*}
where
%\begin{linenomath*}
\begin{align}
		\Cnoz = 83.5\left( \frac{\Lnoz}{\Dnoz}\right)^{-0.22}\left(\frac{\Dnoz}{\Dsac}\right)^{0.15}\,,
		\label{eq:cnoz}
\end{align} 
%\end{linenomath*}
with $\Lnoz$ as the length of the nozzle orifice, $\Dnoz$ as the nozzle exit diameter, and $\Dsac$ as the nozzle sac diameter.
\subsection{Drag Model}\label{sec:drag_model}
The total steady-state drag force on a droplet of diameter $d$ is computed as~\citep{Wan1997,Crowe2012}
%\begin{linenomath*}
\begin{align}
		\wdrag & = \frac{3 \Cdrag \rhog \rhohat \Yhatl(\uhatg-\uhatl)|\uhatg-\uhatl|}{4\rhol \dhat}\,.
		\label{eq:total_drag}
\end{align}
%\end{linenomath*}
The droplet drag coefficient, $\Cdrag$, depends on the droplet Reynolds number, $\Red$, as ~\citep{Wallis1969}
%\begin{linenomath*}
\begin{align}
		\Cdrag = \left\lbrace{\begin{array}{l l} \frac{24}{\Red}(1 + \frac{1}{6}\Red^{2/3}) & \quad \text{for $\Red \leq 1000$ } \\ 0.424 & \quad \text{for $\Red > 1000$} \end{array}} \right.\,,
		\label{eq:cd}
\end{align}
%\end{linenomath*}
with $\Red = \rhog \urel \dhat/\mug$, where $\urel = |\uhatg-\uhatl|$ is the relative velocity and $\mug$ the molecular viscosity of the gas-phase mixture.
\subsection{Droplet Breakup Model}\label{sec:bre_model}
The primary breakup of the liquid jet was not modeled in the original CAS model. Instead, an initial droplet size of $\sqrt{0.1\times\Doo}$ was set at the nozzle exit, after studying sensitivity of liquid penetration. The~\cite{Reitz1986} (\ac{RD}) wave breakup model was used for the secondary breakup, assuming that most breakups were of stripping-type~\citep{Reitz1987} (see~\ref{sec:app_org_model}). In this work, a combined Kelvin-Helmholtz (\ac{KH}) / Rayleigh-Taylor (\ac{RT}) breakup model~\citep{Patterson1998} was incorporated. Under the monodispersed droplet assumption, an equation for the transport of droplet diameter is solved, which is an extension of the well-known $d^2$-law. The droplet breakup source term is computed as $\wbre = \Kbre \rhohat \Yhatl$, where the breakup coefficient, $\Kbre$, is modeled as
%\begin{linenomath*}
\begin{align}
		\Kbre  & = \frac{2\dhat (\dhat-\dst)}{\taub}\,.
		\label{eq:kbre}
\end{align}
%\end{linenomath*}
The stable droplet diameter ($\dst$) and breakup time ($\taub$) are computed from either the KH or the RT breakup model, depending on the local breakup length and time scales of the corresponding models:
%\begin{linenomath*}
\begin{align}
		\dstKH & = 2 B_0 \Lambda_{\mathrm{KH}}\,,
		\label{eq:dst_KH}\\
		\taubKH & = 3.788 B_1 \frac{\dhat}{2 \Lambda_{\mathrm{KH}} \Omega_{\mathrm{KH}}}\,,
		\label{eq:tb_KH}\\
		\dstRT & = C_3 \Lambda_{\mathrm{RT}}\,,
		\label{eq:dst_RT}\\
		\taubRT & = \Omega_{\mathrm{RT}}^{-1}\,.
		\label{eq:tb_RT}
\end{align}
%\end{linenomath*}
Details on the calculation of the growth rate, $\Omega$, and the corresponding wavelength, $\Lambda$, of the fastest growing wave are not given here for brevity and can be found in~\cite{Patterson1998}. The original formulation of the RT model generates multiple droplets at the end of the breakup time. However, in this work, an approach of the continuous breakup is followed for simplicity, which is similar to the KH model and results in a continuous decrease of the droplet diameter. The first, near-nozzle breakup is always modeled by the KH model. Modeling of subsequent breakup events is decided based on the corresponding stable diameter and breakup time from the respective models. If the current droplet diameter is larger than $\dstRT$ and $\taubRT < \taubKH$, the RT model is used; otherwise, the breakup occurs through the KH model. No breakup happens if both $\dstKH$ and $\dstRT$ are larger than the current droplet diameter.\par
For different nozzles, the breakup model constants $C_3$ in \myeq{eq:dst_RT} for the RT model or $\Cb$ in \myeq{eq:old_tb} for the Reitz-Diwakar model need to be tuned to match the experimental liquid length. This is reasonable because the internal nozzle flow can have a significant effect on the primary breakup~\citep{Han2002,Gorokhovski2008,Som2011,Agarwal2019}. The standard ranges of values for the model constants are reported in \mytab{tab:khrt_const}, along with the values used in this work. Although~\cite{Patterson1998} suggested a range of 1.0-5.33 for $C_3$,~\cite{Bravo2014} and~\cite{Wehrfritz2012} have used $C_3$ of 0.1 for evaporating sprays in their simulations. In summary, these constants depend on the nozzle geometry and therefore are tuned for a given injector nozzle. Once tuned for an injector, the model can be used for any operating condition or fuel without any further tuning.
\renewcommand{\arraystretch}{1.0}
\begin{table}[!ht]
\begin{center}
\begin{tabular}{l@{\quad}c@{\quad}r@{\quad}rr}
\toprule 
\textbf{Model} & \textbf{Constant} & \textbf{Standard range} & \multicolumn{2}{r}{\textbf{Present work}} \\ 
\midrule 
Kelvin-Helmholtz & $B_0$ & 0.61 &  \multicolumn{2}{r}{0.61} \\
& $B_1$ & 1.73-60.0 &  \multicolumn{2}{r}{10.0} \\
\midrule
\multirow[t]{4}{*}{Rayleigh-Taylor} & \multirow[t]{4}{*}{$C_3$} & \multirow[t]{4}{*}{0.1-5.33} & Spray A: & 0.60 \\
& & & Spray C:& 0.80 \\
& & & Spray D:& 0.85 \\
& & & FSC:& 0.75 \\
\midrule
\multirow[t]{4}{*}{Reitz-Diwakar} & \multirow[t]{4}{*}{$\Cb$} & \multirow[t]{4}{*}{10.0} & Spray A: & 10.0 \\
& & & Spray C:& 10.0 \\
& & & Spray D:& 12.0 \\
& & & FSC:& 12.0 \\
\bottomrule
\end{tabular}
\end{center}
\caption{Breakup model constants.}
\label{tab:khrt_const}
\end{table}
\subsection{Droplet Size Distribution}\label{subsec:dsd_model}
The mean droplet size along the axis provides little information regarding the distribution of sizes. Early studies of diesel sprays have shown the droplet size distributions to be highly skewed, which can be represented by~\cite{Rosin1933}, logarithmic normal~\citep{Mugele1951},~\cite{Nukiyama1939}, or Chi-square distributions~\citep{Hiroyasu1974}. In recent investigations, lognormal distributions were found to be suitable for diesel sprays~\citep{Feng2019}.\par
We extended the CAS model to consider the droplet size distribution. In this case, the droplet size distribution is modeled as a two-parameter presumed probability distribution function (\ac{PDF}), which can be fitted with parameters depending on moments of droplet diameter. Droplets of all sizes are assumed to be at the local mean droplet temperature ($\Thatd$). The terms dependent on droplet diameters are then integrated over the presumed PDF. The original CAS equation already contained an equation for the second moment of the diameter. Additionally, we derived the equation for the first moment or mean of the droplet diameter. The presumed PDFs are modeled as one of the Rosin-Rammler, lognormal, and gamma distribution functions and used as the boundary condition at the nozzle exit. \mytab{tab:dsd_pdf} shows the PDFs and their moments, which can be used to fit the parameters $m$ and $n$. For lognormal and gamma distributions, closed expressions can be obtained for the parameters as a function of the first and second moments of droplet diameters. However, this is not possible for the Rosin-Rammler distribution, and therefore, the parameters, $m$ and $n$, are computed iteratively. The current limitation of this approach is that the initial droplet size distribution at the nozzle exit must be known either from measurements or primary breakup DNS; otherwise only the delta PDF can be used, which leads to the blob injection model~\citep{Reitz1987,Reitz1987a}.
\renewcommand{\arraystretch}{1.2}
\begin{table}[!ht]
\begin{center}
\begin{tabular}{l@{\qquad}r@{\qquad}r@{\qquad}r}
\toprule
\textbf{PDF} & $ \mathcal{P}(\dhat)$ & $\mean{\dhat}$ & $\mean{\hat{d^2}}$\\ 
\midrule
Rosin-Rammler & $\frac{m}{n}\left( \frac{\dhat}{n}\right)^{m-1} \mathrm{exp}\left(-\left( \frac{\dhat}{n}\right)^{m} \right)$ & $ n \Gamma\left(1+ \frac{1}{m} \right)$ & $n^2 \Gamma\left(1+ \frac{2}{m} \right)$ \\
Log-Normal &$ \frac{1}{\dhat n \sqrt{2\pi}}\mathrm{exp}\left( -\frac{(\mathrm{ln}(\dhat)-m)^2}{2n^2}\right)$ & $\mathrm{exp}\left( m + \frac{n^2}{2}\right)$& $\mathrm{exp}(n^2)\mathrm{exp}\left( 2m +n^2\right)$\\
Gamma & $\frac{n^{m}}{\Gamma(m)} \dhat^{m-1} \mathrm{exp}(-n \dhat)$ & $\frac{m}{n}$& $\frac{m(m+1)}{n^2}$ \\
Delta & $\delta(\dhat'-\dhat)$ & $\dhat$ & $\hat{d^2}$ \\
\bottomrule
\end{tabular}
\end{center}
\caption{Presumed PDFs to model the droplet size distribution.}
\label{tab:dsd_pdf}
\end{table}
\subsection{Transport of Vapor Mass Fraction}
The original CAS model did not track the vapor mass fraction explicitly. Therefore, the far-field vapor mass fraction was assumed to be zero in the evaporation model. While this assumption can be a good approximation for a fast-mixing turbulent spray, it may lead to errors in the calculation of mass transfer number, which may overestimate the evaporation rate and reduce the liquid length. Therefore, an explicit transport equation (\myeq{eq:vap_cont}) is proposed for the vapor mass fraction by splitting the original gas phase continuity equation~\myeq{eq:old_gas_cont}. The source terms, $\wenta b$ and $\wvap b^2$, are mutually exclusive as the entrainment brings purely ambient gas into the spray plume and the evaporation brings pure vapor from the liquid phase into the gas phase. In the gas phase, both ambient gas and fuel vapor are assumed to be perfectly mixed.
\subsection{Evaporation and Droplet Heating Model}\label{sec:evap_model}
The original CAS model used the correlations by \cite{Frossling1938} to model evaporation from the external droplet surface. In this work, we applied the evaporation model of~\cite{Miller1999}, which contains a non-equilibrium treatment of the vapor mass fraction on the droplet surface. The evaporation source term is computed as
%\begin{linenomath*}
\begin{align}
		\wvap & = \frac{3 \Kvap \rhohat \Yhatl}{2\dhat^2}\,,
		\label{eq:wvap}
\end{align} 
where
\begin{align}
		\Kvap &= 4 \frac{\rhog \Gvg}{\rhol}\;\mathrm{ln} (1+\BMd) \Shd\,.
		\label{eq:kvap}
\end{align} 
$\Gvg$ is the diffusion coefficient of fuel vapor (subscript `\ac{v}') in the surrounding gas mixture. The droplet mass transfer number is defined as  
\begin{align}
		\BMd & = \frac{\Ytvs-\Yiv}{1-\Ytvs}\,,
		\label{eq:masstransferno}
\end{align}
where the non-equilibrium vapor mass fraction on the droplet surface is modeled as
\begin{align}
		\Ytvs &  = \frac{\chineqvs}{\chineqvs +  \left(1 - \chineqvs \right) \mathsf{WR}}\,.
		\label{eq:yvs_miller}
\end{align}
where $\mathsf{WR} = W_{\mathrm{a}}/W_{\mathrm{f}}$ is the ratio of the molecular weight of the ambient gas excluding vapor (subscript `\ac{a}') to the molecular weight of fuel (subscript `\ac{f}'). The subscript `\ac{s}' denotes variables at the droplet surface. The non-equilibrium vapor mole fraction,  $\chineqvs$, at the droplet surface is obtained as
\begin{align}
	\chineqvs & =  \chieqvs - \left(\frac{2\LK}{\dhat}\right)\xi\,.
	 \label{eq:chineq}
\end{align}
\cite{Miller1999} proposed the Clausius-Clapeyron relation to obtain equilibrium molar fraction, $\chieqvs$. However, a reasonable calculation of $\chieqvs$ requires appropriate reference values of saturation temperature and pressure. Since the vapor pressure is directly available from the fuel properties, the equilibrium vapor mole fraction, $\chieqvs$, at the droplet surface is computed as
\begin{align}
	\chieqvs & =  \frac{\pv(\Thatd)}{\Pg}\,,
	\label{eq:chieq}
\end{align}
where $\Pg$ is ambient gas pressure and $\pv(\Thatd)$ the vapor pressure at the droplet temperature. The molecular Knudsen layer thickness, $\LK$, is computed as
\begin{align}
	\LK & =  \frac{\mua\left( 2\pi \Thatd \mathcal{R}/\Wv\right)^{1/2}}{\Scg \Pg}\,,
	\label{eq:knudsenlayer}
\end{align}
whereas the non-dimensional evaporation parameter, $\xi$, is given by 
\begin{align}
	\xi & =  \frac{1}{2}\frac{\Prg}{\Scg}\; \mathrm{ln} (1+\BMd) \Shd\,.
	 \label{eq:xi}
\end{align}
%\end{linenomath*}
It is noted that the parameter $\xi$ requires implicit calculations. As reported by~\cite{Miller1999}, $\xi$ is a slowly varying parameter and constant for droplets following the `$d^2$-law'. As a consequence, it is not necessary to perform iterative calculations, and the value from the previous iteration can be used. In this work, the following procedure is used: 
\begin{enumerate}
\item The value of $\BMd$ is computed from the equilibrium $\Ytvs$ using~\myeq{eq:old_yvs},
\item The initial value of $\xi$ is calculated,
\item $\chineqvs$ is computed using~\myeq{eq:chineq} and $\Ytvs$ using~\myeq{eq:yvs_miller},
\item $\BMd$ is then updated using the non-equilibrium $\Ytvs$,
\item $\xi$ is updated.
\end{enumerate}
The far-field vapor mass fraction, $\Yiv$, in the calculation of the mass transfer number~\myeq{eq:masstransferno} is set according to the one-third rule~\citep{Hubbard1975} to $\Yiv = \Cvap\,(\Yhatv +2 \Ytvs)/3$. An additional scaling constant, $\Cvap = b(0,t)/b(z,t)$, is introduced to account for the effects of locally entrained fresh ambient gas encountered by droplets away from the centerline. Differently, in the original CAS model, $\Yhatv$ was set to zero assuming the far-field value of the vapor mass fraction to be negligible.\par
The droplet temperature is affected by continuous heating and evaporation. The respective source term is modeled as
$\wheat = \Kheat \rhohat \Yhatl$ with the temperature coefficient, $\Kheat$, given by
%\begin{linenomath*}
\begin{align}
		\Kheat  = \frac{6\Qd}{\rhol \dhat \Cl(\Thatd)} - \frac{3 \Kvap L(\Thatd)}{2 \dhat^2 \Cl(\Thatd)}\,,
		\label{eq:kt}
\end{align}
where $L(\Thatd)$ and $\Cl(\Thatd)$ are the latent heat of vaporization and the heat capacity of the fuel at the droplet temperature $\Thatd$, respectively. The heat transfer between the droplet and ambient gas mixture is modeled as
\begin{align}
	\Qd = \frac{\lambdag(\Tref) (\Thatg - \Thatd)}{\dhat} f_2 \Nud\,,
	\label{eq:qd_miller}
\end{align}
where $\lambdag(\Tref)$ is the thermal conductivity of the local gas-phase mixture and $\Thatg$ the ambient gas temperature. Since the vapor mass fraction is now known, $\Thatg$ can be computed assuming homogeneous mixing of vapor in the ambient gas as
{\color{black}
\begin{align}
		\Thatg = \frac{\Yhatv \Thatv + \Yhata \Thata}{\Yhatg},
		\label{eq:Tgasnew}
\end{align}
where the vapor temperature $\Thatv$ is calculated iteratively from the droplet temperature by using the latent heat of vaporization for evaporation.}
The analytical evaporative heat transfer correction, $f_2$, is given by
\begin{align}
	f_2 = \frac{\xi}{e^{\xi}-1}\,.
	\label{eq:f2}
\end{align}
For completeness, the correlations for the droplet Sherwood and Nusselt numbers from~\cite{Ranz1952} are given here:
\begin{align}
	\Shd & = 2 + 0.552~\Red^{1/2}~\Scg^{1/3}\,,
	 \label{eq:sherwoodno_miller}\\
	\Nud & = 2 + 0.552~\Red^{1/2}~\Prg^{1/3}\,.
	\label{eq:nusseltno_miller}
\end{align}
The gas-phase Schmidt number, $\Scg$, is defined as
\begin{align}
\Scg = \frac{\mug}{\rhog \Gvg}
\end{align}
and the gas-phase Prandtl number, $\Prg$, is defined as
\begin{align}
\Prg = \frac{\mug(\Tref) \Cpg(\Tref)}{\lambdag(\Tref)}\,.
\label{eq:prg}
\end{align}
$\Cpg(\Tref)$ is the specific heat capacity of the gas-phase mixture at constant pressure, which is computed by the linear mixing rule as
\begin{align}
		\Cpg = \frac{\Yhatv \Cpv + \Yhata \Cpa}{\Yhatg}\,.
		\label{eq:cpg}
\end{align}
The gas-phase mixture properties and correlations are computed at the reference temperature, which is obtained by the one-third rule~\citep{Hubbard1975} as
\begin{align}
	\Tref = \frac{\Thatg+2\Thatd}{3}\,.
	\label{eq:tref}
\end{align}
%\end{linenomath*}
Since the boiling model is not used, the droplet temperature is kept bounded by an upper limit of saturation temperature at the given pressure. The liquid properties such as density ($\rhol$), surface tension ($\sigmal$), and viscosity ($\mul$) are evaluated at a constant fuel temperature at the nozzle exit ($\Tlo$) and remain constant throughout the liquid phase. The gas-phase mixture properties, such as viscosity ($\mug$) and thermal conductivity ($\lambdag$), are computed from the vapor and ambient gas properties by~\cite{Wilke1950} formula.
\subsection{Nozzle Flow Model}\label{subsec:nozzle_flow_model}
In the original CAS model, injection velocities were computed using simple Bernoulli relations considering the measured discharge coefficient ($\Cd$). In this work, a phenomenological zero-dimensional nozzle flow model from~\cite{Sarre1999} is used to predict the nozzle exit velocity. This model considers the influence of nozzle geometry, the development of separation inside the nozzle, possible cavitation, as well as the hydraulic flip phenomenon. The initial nozzle exit velocity is computed from Bernoulli's equation, assuming the discharge coefficient of unity. The discharge coefficient is then computed iteratively using a correlation based on inlet and expansion losses. The final mean exit velocity is computed from the converged value of the discharge coefficient. The nozzle flow model provides the effective jet velocity and effective jet diameter, which can be different from the geometric nozzle exit diameter if cavitation or hydraulic flip occur within the nozzle.
\subsection{Numerical Methods}
The hyperbolic system of PDEs~\myeq{eq:amb_cont}-\eqref{eq:liq_energy} is numerically solved in conservative form. An upwind convective scheme is required to account for the possible discontinuities in the multiphase system.~\cite{Wan1997} used a combination of~\cite{Beam1976} and~\cite{MacCormack1969} schemes to avoid oscillations in the solution. In this work, various numerical schemes such as first order Lax-Friedrichs~\citep{Lax1954} and the third and fifth order weighted essentially non-oscillatory (\ac{WENO}) schemes~\citep{Jiang1996} are implemented. Numerical tests did not show significant benefit in the accuracy of quantities of interest with higher order schemes. Therefore, the Lax-Friedrichs scheme employing Rusanov fluxes~\citep{Rusanov1961} with local wave speeds is chosen to keep computational costs low. An explicit Euler scheme is used for the time advancement. Since different physical processes, which are lumped together in the source terms, have different time scales, a Courant-Friedrichs-Lewy (\ac{CFL}) number of 0.1-0.5 is used to ensure stability. The governing equations and models are implemented in an in-house FORTRAN code framework, named CARTS.

\subsection{Solution Procedure}
The variables in the governing equation are non-dimensionalized using the effective jet diameter $\Do$ as the length scale, the effective jet velocity $\Uo$ computed using nozzle flow model in \mysubsec{subsec:nozzle_flow_model} as the velocity scale, and $\tauo=\Do / \Uo$ as the time scale. The mixture density and temperatures are non-dimensionalized by the liquid phase density ($\rhol$) and nozzle exit temperature ($\Tlo$), respectively. The non-dimensional (\ac{ND}) system of equations is discretized in space and time, the variables are initialized in the 1D discretized domain, and boundary conditions (\ac{BC}s) are applied with the left boundary as the liquid jet and the right boundary treated as the far-field condition. The initial (\ac{IC}) and boundary conditions of the ND variables are given in~\mytab{tab:init_bc}. 
\renewcommand{\arraystretch}{1.0}
\begin{table}[!b]
\begin{center}
\begin{tabular}{l@{\quad}l@{\quad}c@{\quad}l@{\quad}r@{\quad}r}
\toprule
\textbf{Variable} & \textbf{Definition} & \textbf{IC} & \textbf{Left BC} & \textbf{Right BC} & \textbf{Bounds} \\
\midrule
$\Ystarl$ & $\Yhatl$ & 0.0 & Dirichlet 1.0 & Neumann 0.0 & [0.0,1.0] \\
$\Ystarv$ & $\Yhatv$ & 0.0 & Dirichlet 0.0 & Neumann 0.0 & [0.0,1.0] \\
$\Ystara$ & $\Yhata$ & 1.0 & Dirichlet 0.0 & Neumann 0.0 & [0.0,1.0] \\
$\rhostar$ & $\rhohat/\rhol$ & $\rhoa/\rhol$ & Dirichlet 1.0 & Neumann 0.0 & [$\rhog/\rhol$,1.0] \\
$\ustarl$ & $\uhatl/\Uo$ & 0.0 & Dirichlet 1.0$^{\star}$ & Neumann 0.0 & [0.0,1.0] \\
$\ustarg$ & $\uhatg/\Uo$ & 0.0 & Dirichlet 0.0 & Neumann 0.0 & [0.0,1.0] \\
$\mean{\dstarsq}$ & $\mean{\dhat^2}/\Doo$ & 0.0 & Dirichlet 0.1-1.0 & Neumann 0.0 & [0.0,1.0] \\
$\mean{\dstar}$ & $\mean{\dhat}/\Do$ & 0.0 & Dirichlet 0.1-1.0 & Neumann 0.0 & [0.0,1.0] \\
$\Tstard$ & $\Thatd/\Tlo$ & 0.0 & Dirichlet 1.0 & Neumann 0.0 & [0.0,$\Tsat/\Tlo$] \\
$\Tstarg$ & $\Thatg/\Tlo$ & $\Tamb/\Tlo$ & Neumann 0.0 & Neumann 0.0 & [0.0,$\Tamb/\Tlo$] \\
$\bstar$ & $b/\Do$ & 0.5 & Dirichlet 0.5 & Neumann 0.0 & [0.5,$\infty$) \\
$\zstar$ & $z/\Do$ & - & 0.0 & 1600.0 & [0.0,1600.0] \\
$\tstar$ & $t/\tauo$ & 0.0 & - & - & [0.0,$\infty$) \\
\bottomrule
\end{tabular}
\end{center}
\caption{Initial and boundary conditions.$^{\star}$Normalized rate of injection profile is applied, if available.}
\label{tab:init_bc}
\end{table}
First, the conservative ND variables are advanced in time. The primitive ND variables are then computed from the conservative ND variables. Numerical difficulties occur when liquid- or gas-phase mass fractions become small. A threshold of $10^{-6}$ for the liquid phase mass fraction, $\Ystarl$, produces a numerically stable solution. Below this threshold, all liquid phase variables are set to zero. In addition, the summation of mass fractions is ensured to be unity, adjusting the value of the ambient gas mass fraction ($\Ystara$). The primitive variables are bounded by their physical limits to avoid an unphysical solution. For the cases with droplet size distribution, the droplet diameters are discretized into a fixed number (in this work, 200) of classes about the mean diameter to represent the PDF.
%\begin{align}
%		\ustarlg = \frac{\uhatlg}{\Uo},~\zstar =\frac{z}{\Do},~\tstar =\frac{t}{\tauo},~\bstar = \frac{b}{\Do},~\dstar = \frac{\dhat}{\Do},~\rhostar = \frac{\rhohat}{\rhol},~\Tstard = \frac{\Thatd}{\Tlo},~\Tstarg = \frac{\Thatg}{\Tlo},~\Ystarlg = \Yhatlg
%		\label{eq:nd}
%\end{align}

\section{Description of Cases and Validation Data}\label{sec:case_desc}
\subsection{ECN Data}
Selected experimental data from the Engine Combustion Network~\cite{ECN2020} are used for validation of the model under a wide range of operation conditions. The initial and boundary conditions corresponding to the measurement data of the selected inert inert cases are summarized in~\mytab{tab:cases} (cases 1-11). A detailed description of the cases and the measurement setup can be found on the \ac{ECN} webpage~\citep{ECN2020} and in previous literature~\citep{Pickett2010,Pickett2011,Skeen2015}. \mytab{tab:noz_geo} shows the geometric details and the hydraulic coefficients of the nozzles considered in this work. The rate of injection is generated using the virtual injection rate generator~\citep{VIR2020}.
\renewcommand{\arraystretch}{1.0}
\begin{table}[h]
\begin{center}
\begin{tabular}{r@{\quad}l@{\quad}l@{\quad}r@{\quad}r@{\quad}r@{\quad}r@{\quad}r@{\quad}r@{\quad}r}
\toprule
\textbf{No.} $\downarrow$ & \textbf{Nozzle} & \textbf{Fuel} & $\Pinj$ & $\Tinj$ & $\Pamb$ & $\Tamb$ & \textbf{DOI} & $\theta\tsup{*}$ & $\theta\tsup{**}$  \\ 
\midrule 
Unit $\rightarrow$& - & - & \si{\bar} & \si{\kelvin} & \si{\bar} & \si{\kelvin} & \si{\milli\second} & \si{\degree} & \si{\degree} \\
\midrule 
1 & Spray A & \ndo & 1500 & 373 & 60 & 900 & 1.5 & - & 15 \\
2 & Spray A & \ndo & 1500 & 373 & 40 & 900 & 1.5 & - & 13.6 \\
3 & Spray A & \ndo & 1500 & 373 & 20 & 900 & 1.5 & - & 11.4 \\
4 & Spray A & \ndo & 1500 & 373 & 10 & 900 & 1.5 & - & 9.4 \\
5 & Spray A & \ndo & 1000 & 373 & 60 & 900 & 1.5 & - & 15 \\
6 & Spray A & \ndo & 500 & 373 & 60 & 900 & 1.5 & - & 15 \\
7 & Spray A & \ndo & 1500 & 363 & 30 & 440 & 1.5 & - & 15 \\
8 & Spray A & \ndo & 1500 & 373 & 46 & 700 & 1.5 & - & 15 \\
9 & Spray A & \ndo & 1500 & 403 & 80 & 1200 & 1.5 & - & 15.2 \\
10 & Spray C & \ndo & 1500 & 373 & 60 & 900 & 1.5 & - & 20.0 \\
11 & Spray D & \ndo & 1500 & 373 & 60 & 900 & 1.5 & - & 16.0 \\
12 & FSC & \oct & 1400 & 353 & 50 & 835 & 1.5 & 16.4 & 18.2 \\
13 & FSC & \oct & 900 & 353 & 34 & 835 & 1.5 & 14.2 & 16.4 \\
14 & FSC & \dnbe & 1400 & 353 & 50 & 835 & 1.5 & 17 & 18.6 \\
15 & FSC & \dnbe & 900 & 353 & 34 & 835 & 1.5 & 15 & 16.8 \\
\bottomrule
\end{tabular}
\end{center}
\caption{Cases chosen for validation in this study. $\tsup{*}$Measured vapor cone angle. $\tsup{**}$Computed by~\myeq{eq:theta}. Breakup model constants $\Cb$ and $C_3$ were tuned for injector nozzles for cases 1, 10, 11, and 12.}
\label{tab:cases}
\end{table}
\renewcommand{\arraystretch}{1.0}
\begin{table}[h]
\begin{center}
\begin{tabular}{p{0.21\textwidth}@{\quad}p{0.1\textwidth}@{\quad}p{0.05\textwidth}@{\quad}>{\raggedleft}p{0.11\textwidth}@{\quad}>{\raggedleft}p{0.11\textwidth}@{\quad}>{\raggedleft}p{0.11\textwidth}@{\quad}p{0.09\textwidth}<{\raggedleft}}
\toprule 
 &  &  & \multicolumn{4}{c}{\textbf{Nozzles}} \\ \cline{4-7}
\multirow{-2}{*}{\textbf{Parameter}} & \multirow{-2}{*}{\textbf{Symbol}} & \multirow{-2}{*}{\textbf{Unit}} & \textbf{Spray A} &   \textbf{Spray C} & \textbf{Spray D} & \textbf{FSC} \\ 
\midrule 
\# of holes & - & - & 1 & 1 & 1 & 8 \\ 
Sac diameter & $\Dsac$ & \si{\micro\metre} & 660 & 660 & 2500* & 124 \\ 
Exit diameter & $\Dnoz$ & \si{\micro\metre} & 90 & 200 & 189 & 109 \\ 
Inlet radius to exit diameter ratio & $\rnoz / \Dnoz$ & - & 0.256 & 0.2 & 0.55 & 0.125 \\ 
Length to exit diameter ratio & $\Lnoz/\Dnoz$ & - & 11 & 5 & 5.5 & 9.4 \\ 
Inlet loss coefficient & $K_\mathrm{{in}}$ & - & 0.035 & 0.039 & 0.039 & 0.085 \\ 
Expansion loss coefficient & $K_\mathrm{{exp}}$ & - & 0.0 & 0.0 & 0.0 & 0.01 \\ 
Discharge coefficient (measured) & $\Cd$ & - & 0.89 & - & - & 0.86-0.91 \\ 
Discharge coefficient (computed) & $\Cd$ & - & 0.89 & 0.95 & 0.94 &  0.87-0.90 \\ 
\bottomrule
\end{tabular} 
\end{center}
\caption{Geometric and hydraulic parameters for different nozzles. *Sac diameter for Spray D is increased from 660\,$\mu m$ to obtain a reasonable cone angle through the model.}
\label{tab:noz_geo}
\end{table}
\subsection{FSC  Data}
In addition to the ECN data, the data generated within the \ac{FSC} are used (cases 12-15) to evaluate the model for the effect of different fuel properties. The data have been measured in a pressure chamber, which is built as a heated constant-pressure optically accessible flow vessel (ambient gas velocity, $u_\mathrm{a}$ $\leq$ \SI{0.1}{\metre/\second}). Ambient conditions up to \SI{835}{\kelvin} and \SI{50}{\bar} are presented. The 8-hole injector is centrally positioned in the chamber so that one of the spray plumes propagates vertically from bottom to top, which is then investigated. The fuel is provided by a common-rail injection system with a maximum investigated injection pressure of \SI{1400}{\bar}.\par
For measuring the liquid penetration, Mie scattering visualizations were used, while the Schlieren technique was used for determining vapor penetration. The optical access of the pressure vessel limits the maximum detectable spray penetration to \SI{55}{\milli\metre}. To determine the vapor penetration and cone angles in the Schlieren images, the spray must be separated from the surrounding gas phase to guarantee a valid binarization. A Gaussian Fast Fourier Transformation Filter is used to suppress the diﬀerent textures/frequencies of the surrounding ambience. Afterwards, the post-processing procedure for the Mie-scattering based on background subtraction and a fixed intensity threshold for binarization can be applied to the filtered Schlieren images. Only the first half of the spray is considered to measure the steady-state cone angles, which are not influenced by entrained air~\citep{Naber1996}. Fuel injections were repeated 20 times with a frequency of \SI{0.2}{\hertz}. A detailed description of the performed measurements can be found in~\cite{Palmer2015}.
%\begin{figure}[h]
%	\centering
%	\includegraphics[scale=1.0]{./figures/fig1_OpticalSetup.eps}
%	\caption{Schematic illustration of the Mie-scattering (top) and the Schlieren setup (bottom)~\citep{Palmer2015}.}
%	\label{fig:exp}
%\end{figure}
\par
The measured vapor cone angles and the spray cone angles computed by~\myeq{eq:theta} are reported in~\mytab{tab:cases} for comparison. The rate of injection was not measured. Therefore, appropriate linearly increasing rate of injection proﬁles were used to closely reproduce initial stages of the spray evolution as observed in the experiments.
\subsection{Simulation Data}
Additional validation is performed with the results of a 3D LES as well as unsteady RANS (\ac{URANS}) simulations as reference and benchmark for the 1D model. The LES was performed using an in-house code called CIAO, which solves the Navier-Stokes equations along with multiphysics effects. CIAO is a structured, arbitrary order, finite difference code~\citep{Desjardins2008,Mittal2014}. Lagrangian particle tracking was used for the discrete liquid phase. Simulations were performed using a fourth order scheme for the density and momentum equations and WENO5 scheme for the scalars, on the stretched grid with a minimum grid spacing of \SI{80}{\micro\meter} at the nozzle orifice and maximum grid spacing of approximately \SI{600}{\micro\meter} at the farthest location downstream. The combined Kelvin-Helmholtz Rayleigh-Taylor (\ac{KHRT})~\citep{Patterson1998} breakup model and ~\cite{Miller1999} evaporation models were used in the simulation. For details on methods, models, and numerical setup, the reader is referred to~\cite{Davidovic2017}.\par
The commercial 3D CFD software CONVERGE from Convergent Science Inc.~\citep{Richards2017} was employed for the URANS simulation of the Spray A case under inert conditions. The Spray A case has been widely studied using CONVERGE~\citep{Senecal2014,Wang2014}. The URANS simulations in this work used the base grid size of \SI{1.5}{\milli\meter} with near-nozzle fixed embedding to level 3, resulting in the smallest grid size of \SI{0.1875}{\milli\meter}. Adaptive mesh refinement (\ac{AMR}) up to level 3 was used outside the fixed embedding, which enabled local refinement depending on the flow conditions. This resulted in the cell count varying in the range of \numrange{3e5}{1.2e6}. The KH-RT breakup model and~\cite{Frossling1938} evaporation model were used in the simulation.\par
The differences in physical models and parameters between the 3D CFD model and the 1D CAS model are noted. Simpler physical models in the CAS model reduce the computational costs and therefore are preferred. For example, while URANS used the dynamic drag model, the spherical drag model was used in the LES and CAS model. Similarly, the collision model was not used in the LES and CAS model. In 3D CFD, droplet breakup and evaporation result in droplets of various sizes at each axial location, whereas in the CAS model, the droplets are either assumed to be monodispersed across the cross-section or presumed PDFs are used to represent the local droplet size distribution as described in \mysubsec{subsec:dsd_model}. Due to Lagrangian-Eulerian approach in the 3D simulations, the mixture density does not contain contributions from the liquid phase. Therefore, the density-weighted cross-sectional averages of the available flow variables were computed differently compared to those defined by \myeq{eq:radint}. The boundaries of the spray plume were first determined by the threshold vapor mass fraction of 0.035. Within the spray boundaries, the averages of the flow variables were then computed. The threshold value was entered by a visual inspection of the spray plume, as lower values resulted in excessive dilution of the flow variables by their far-field values.
%\begin{figure}[!ht]
%	\centering
%	\includegraphics[scale=0.4]{./figures/roi_profile}
%	\caption{Rate of injection profile for different injectors.}
%	\label{fig:roi}
%\end{figure}

\section{Results and Discussion}~\label{sec:results}
The solution of the governing equations of the CAS model provides flow variables such as density, mass fractions and velocities of the liquid and vapor phases, mean droplet diameter, and the droplet temperature along the axial co-ordinate. From these data, the global spray characteristics such as liquid and vapor penetration were extracted. The ECN guidelines suggest the liquid penetration length to be determined as the farthest axial location from the nozzle exit where the liquid volume fraction is \SI{0.1}{\percent}, and the vapor penetration length to be determined as the farthest axial location from the nozzle exit where the vapor mass fraction is \SI{0.1}{\percent}. Since the vapor mass fraction is not explicitly tracked in the original CAS model, the vapor or gas penetration is determined as the axial location from the nozzle exit where the gas velocity normalized by the nozzle exit velocity ($\Uo$) is \SI{0.1}{\percent}. This definition was found to be equivalent to that of the ECN. First, the original CAS model was evaluated for a wide range of conditions and fuels. The incremental improvements in the model performance are then discussed, as the new sub-models are added or replaced.
\subsection{Grid Convergence}
\begin{figure}[h]
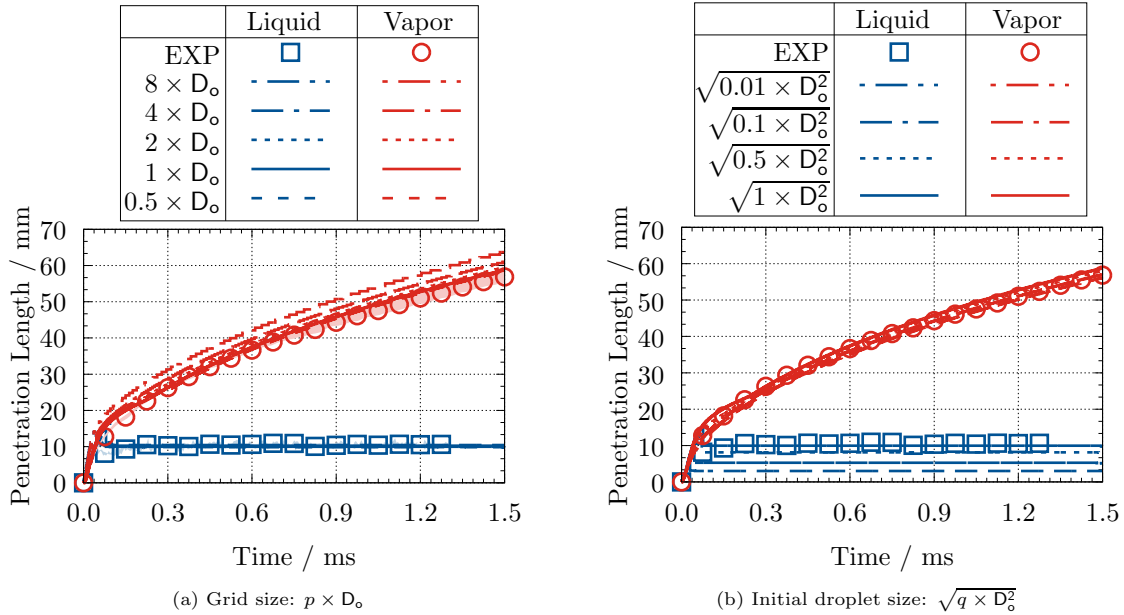

	\begin{subfigure}[b]{0.49\textwidth}
	\centering
	\input{./figures/OriginalCasGridConvergence.tex}
	\caption{Grid size: $p\times \Do$}
	\label{fig:gridcnvg}
	\end{subfigure}
	~
	\begin{subfigure}[b]{0.49\textwidth}
	\centering
	\input{./figures/OriginalCasInitDropletSize.tex}
	\caption{Initial droplet size: $\sqrt{q \times \Doo}$}
	\label{fig:initdropsize}
	\end{subfigure}
	\caption{Effect of the grid size ($p=0.5,1,2,4,8$) and the initial droplet size ($q=0.01,0.1,0.5,1.0$) on the liquid and vapor penetration lengths tested on case 1.}
	\label{fig:gridinit}
\end{figure}

A grid convergence study was performed with the original CAS model to select the grid size for further computations. \myfig{fig:gridcnvg} shows the effect of increasing grid resolution on the liquid and vapor penetration for the Spray A case 1. The grid resolution is represented in terms of the effective jet diameter (d$z$ = $p\times \Do$). Both liquid and vapor penetration converge as the grid size is decreased. From this study, the grid size of $1 \times \Do$ was chosen as a compromise between accuracy and computational cost, and the simulations reported in the following subsections were performed with this chosen grid size. 
\subsection{Initial Droplet Size}
In the original CAS model, the initial droplet size of $\sqrt{0.1\times \Doo}$ was suggested. In this work, various initial droplet sizes were tested to evaluate their effect on the liquid and vapor penetration lengths. Results in \myfig{fig:initdropsize} (for Spray A case 1) show a significant effect of the initial droplet size on the liquid penetration, whereas the vapor penetration is hardly affected, except in the initial phase before \SI{0.2}{\milli\second}. The sensitivity to the initial droplet size was also verified for other selected cases, but results are not shown here for brevity. The initial droplet size of $\sqrt{1\times \Doo}$ can predict the liquid and vapor penetration reasonably well, and therefore it was used for all cases in this work. This corresponds to the blob injection model~\citep{Reitz1987,Reitz1987a}.
\subsection{Results with the Original CAS Model}
\begin{figure}[h]
	\centering
	\input{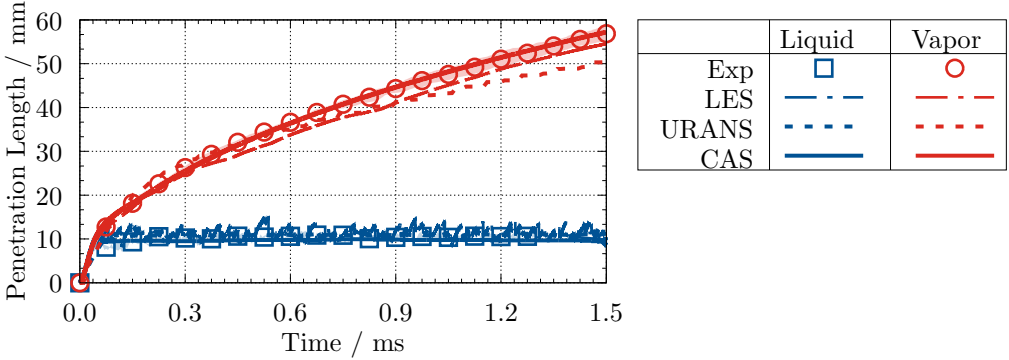}
	\caption{Comparison of the liquid and vapor penetration lengths in the LES, the URANS simulation, and the CAS model for case 1.}
	\label{fig:org_cas_validation}
\end{figure}

For validation, the liquid and vapor penetration lengths for case 1 with inert ambient conditions are compared in~\myfig{fig:org_cas_validation}. The liquid penetration length from CAS agrees well with URANS, LES, and the experimental values. The vapor penetration length from the CAS model agrees well with the measurements. While LES predicts the vapor penetration quite well, the URANS results deviate from the experiments after approximately \SI{0.7}{\milli\second}. Such deviations have also been observed in~\cite{Senecal2014} in the grid-converged results. The results of the URANS simulations performed in this work are close to the converged values with the finer resolution in their grid convergence study. Despite model reductions, the 1D CAS model captures the vapor length closely. The reasons for such agreement can be attributed to the strong dependency on the initial
momentum and correct momentum transfer from the liquid to the gas phase due to the Eulerian description. Previous studies have shown that incorrect coupling between the liquid and gas phases can occur in Lagrangian particle tracking methods, and thus spray penetration can become highly grid-dependent~\citep{Abraham1997,Beard2000,Post2000,Qiu2015,Wei2017}. In the recent studies, Eulerian spray models have been able to correctly reproduce the vapor penetration in URANS~\citep{Garcia2013,Blanco2016} and LES~\citep{Matheis2018,Desantes2020}. Another factor affecting vapor penetration is the turbulence model, which has an influence on the entrainment of ambient gas. Studies have shown the RANS models to underpredict the vapor penetration at longer times, while the LES results always matched the measurements well~\citep{Zhou2011,Xue2013,Kahila2018}. The entrainment of the ambient gas in the CAS model comes through the spreading of the spray specified by the spray angle and implicitly considers the turbulence effects.
\par
\begin{figure}[]
	\centering
	\input{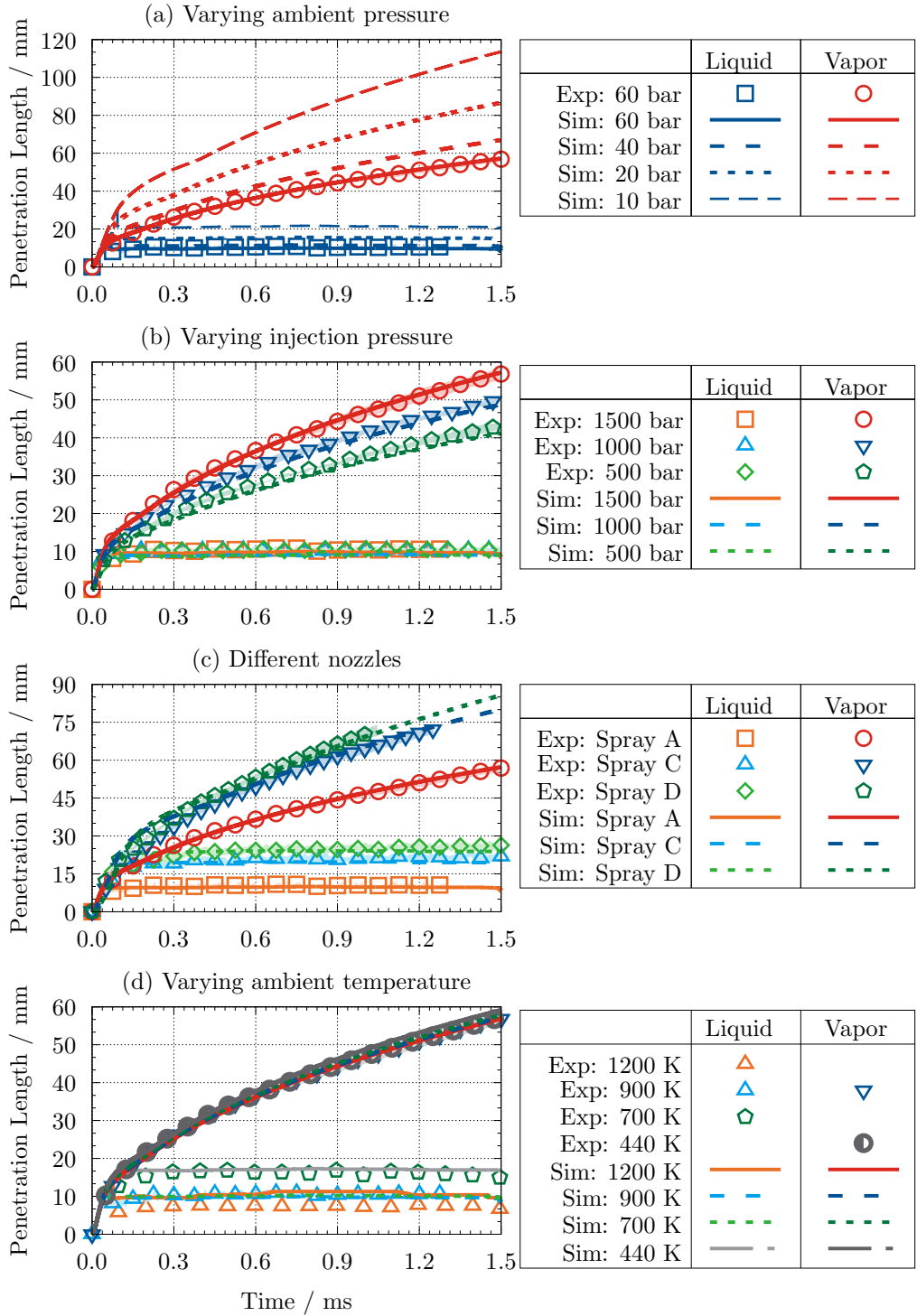}
	\caption{Effect of ambient pressure, injection pressure, nozzle geometry, and ambient temperature on the liquid and vapor penetration lengths (cases 1-11).}
	\label{fig:org_cas_all}
\end{figure}
The original CAS model is then applied to ECN cases (2-11) from \mytab{tab:cases}, and the performance of the model is analyzed with respect to varying ambient pressures (or densities), injection pressures, ambient temperatures, and nozzles. Experimental data are shown where available. The effects of ambient pressure (or density) on spray penetration are well-known from previous studies~\citep{Naber1995,Siebers1998,Siebers1999}, where a strongly non-linear trend of the liquid and vapor lengths with respect to ambient pressure has been reported. This effect is reproduced well by the model, as shown in \myfig{fig:org_cas_all}(a). The variation of injection pressure has a primary effect on the vapor penetration but no observable effect on the liquid penetration (see \myfig{fig:org_cas_all}(b)), which is also captured well by the model. The injection pressure directly controls the momentum flux at the nozzle exit, which has an impact on the vapor penetration, or in other words, the spray tip penetration. On the other hand, the liquid penetration is governed by breakup and evaporation processes, which are not significantly changing as the injection pressures are sufficiently high.\par
\myfig{fig:org_cas_all}(c) shows the effect of different nozzles. For different nozzles, the breakup model constant $\Cb$ in \myeq{eq:old_tb} needed to be tuned to match the experimental liquid length as discussed in \mysec{sec:bre_model}. It includes the effects of the internal nozzle flow on the primary breakup. While the recommended value of 10.0 for $\Cb$ was used for Spray A and Spray C, it was increased to 12.0 for Spray D. For these cases, the results cannot be considered as validation. However, it must be noted that the values of model constants are calibrated only once for a given injector. Once tuned, the model should be expected to produce reasonable trends in liquid and vapor lengths for a wide range of operating conditions and fuels.\par
\myfig{fig:org_cas_all}(d) shows the effect of varying ambient temperatures on spray characteristics. The vapor penetration is affected less by ambient gas temperatures compared to the liquid penetration. On the other hand, the liquid length decreases with increasing gas temperature as observed in the measurements. The model is not able to predict these trends in liquid length with respect to the gas temperature.\par
\begin{figure}[!t]
	\centering
	\input{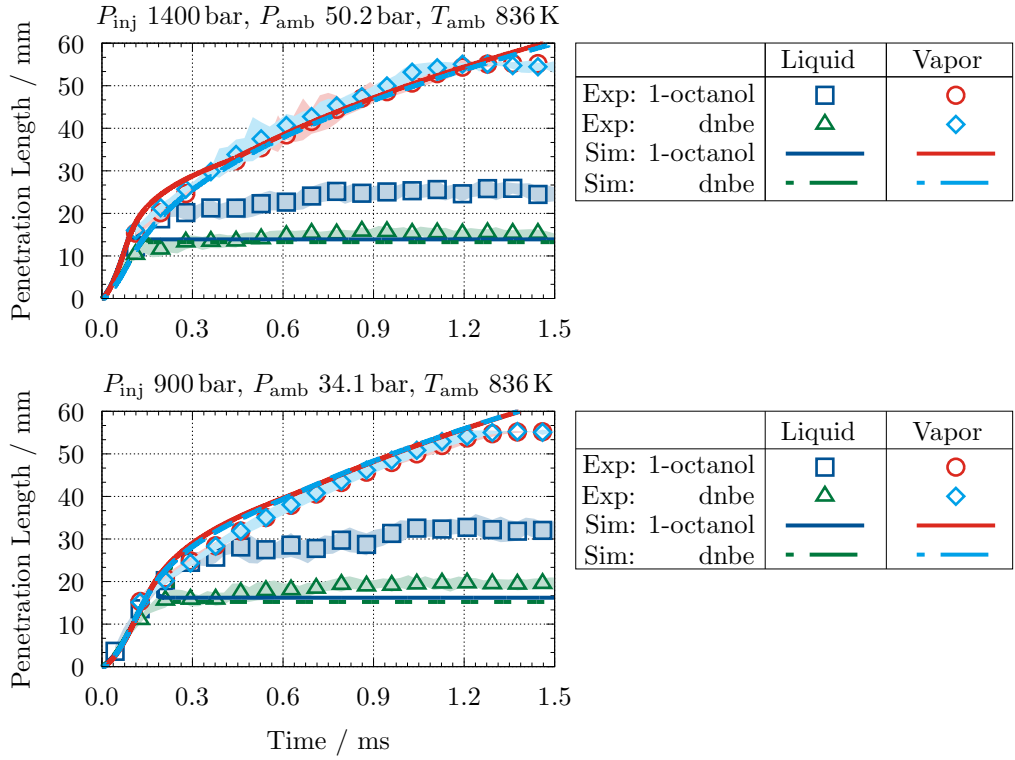}
	\caption{Effect of different fuel properties on the liquid and vapor penetration lengths (cases 12-15).}
	\label{fig:org_cas_fuel}
\end{figure}

The model is then applied to FSC cases (12-15) with two different fuels, namely \oct~and \dnbe~(dnbe), and two different ambient conditions (see~\myfig{fig:org_cas_fuel}). For both conditions, the liquid length for \oct~is higher because of higher heat of vaporization (see \myfig{fig:fuelprop}(a)) and lower vapor pressure (see \myfig{fig:fuelprop}(b)) in comparison with \dnbe. The model is not able to bring these differences in the fuel properties into the liquid lengths of the two fuels. Similar to Spray D, the value of $\Cb$ was set to 12.0 for the FSC nozzle. 
\begin{figure}[!ht]
	\centering
	\input{./figures/FuelProperties.tex}
	\caption{Selected properties of \oct~and \dnbe~\citep{Daubert1989}.}
	\label{fig:fuelprop}
\end{figure}
\subsection{Results with Model Improvements}\label{sec:improvements}
The CAS model, in its original formulation, can reasonably predict the effects of ambient pressure, injection pressure, and nozzle geometry on liquid and vapor penetration of sprays. However, it fails to capture the effect of ambient temperature and, more importantly, the effect of fuel properties, as discussed in the previous section. Therefore, the model was improved in its predictive capabilities by updating the current standard breakup and evaporation models, as described in \mysec{sec:new_model}. In addition, a transport equation of vapor mass fraction was added.
\subsubsection{KH-RT Breakup Model}
\begin{figure}[!t]
	\centering
	\input{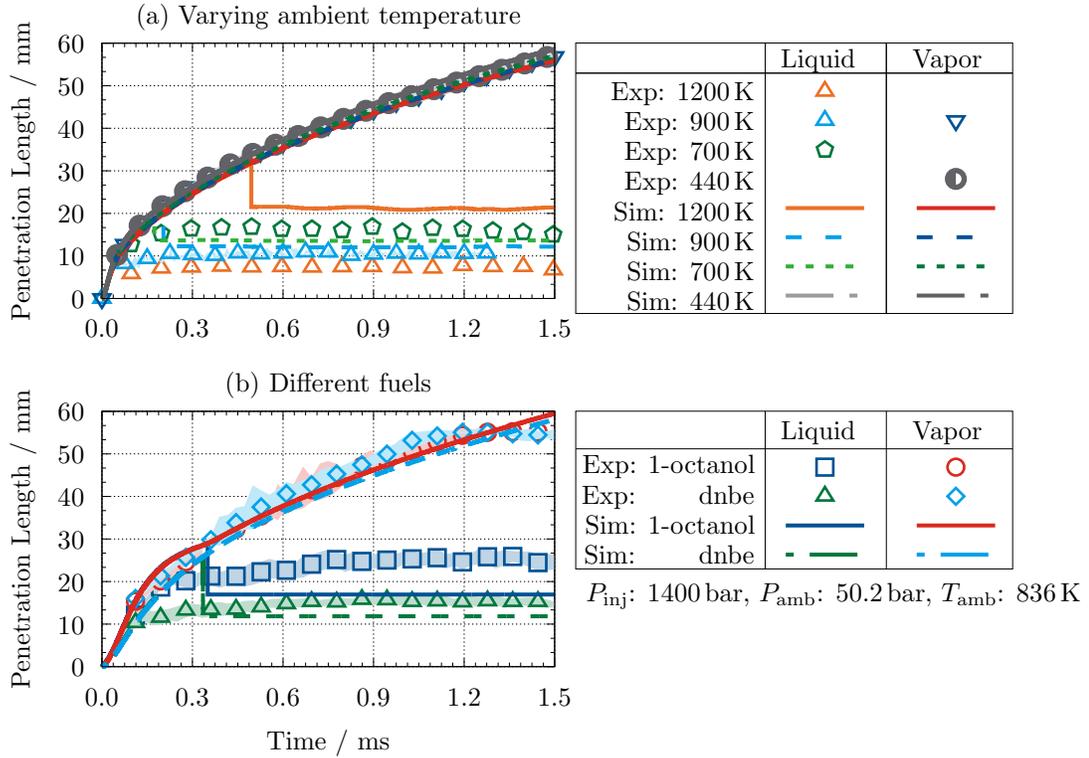}
	\caption{Results using the KH-RT breakup model for (a) cases 1 and 7-9 and (b) cases 12 and 14.}
	\label{fig:khrt_bre}
\end{figure}

The simulations for all cases (1-15) were performed with the combined KH-RT breakup model. For brevity, the results are reported here only for the cases showing significant deviations from the trends. \myfig{fig:khrt_bre}(a) shows the results for varying ambient temperatures. With increasing ambient temperature, the liquid penetration should decrease as observed in the experiments. This trend is followed for temperatures from \SIrange{440}{900}{\kelvin}. For the temperature of \SI{440}{\kelvin}, there is no clear distinction between liquid and vapor penetration, which is consistent with the measurements. However, for the temperature of \SI{1200}{\kelvin}, the predicted liquid penetration is not consistent with the trend and measurements.\par
Compared to the Reitz-Diwakar breakup model, the combined KH-RT breakup model captures the effects of different fuel properties to a certain extent, as shown in \myfig{fig:khrt_bre}(b). However, the predicted liquid lengths are lower than those measured for both fuels.
\subsubsection{Transport of Vapor Mass Fraction}
\begin{figure}[!ht]
	\centering
	\input{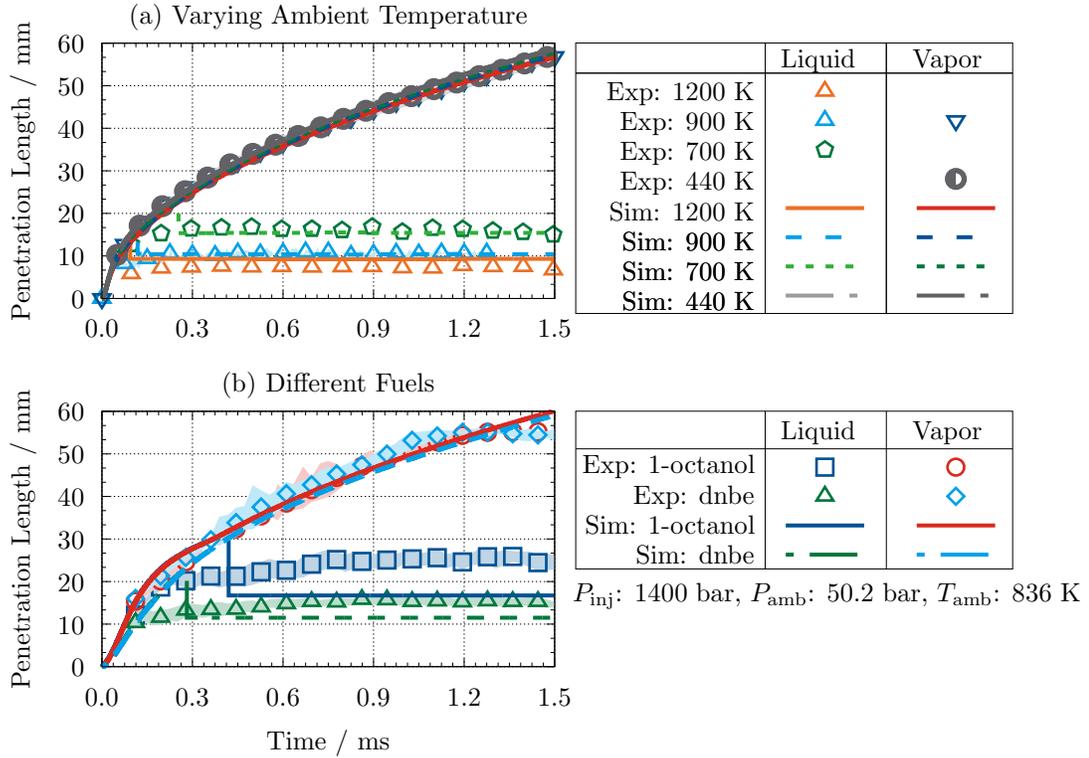}
	\caption{Results using the KH-RT breakup model along with the additional transport equation for the vapor mass fraction for (a) cases 1 and 7-9 and (b) cases 12 and 14.}
	\label{fig:yv_transport}
\end{figure}
\myfig{fig:yv_transport} shows the results for varying ambient temperatures and different fuels with the additional information on the vapor mass fraction. The trends in the temperature variation are improved significantly, particularly for the case at the temperature of \SI{1200}{\kelvin}. However, the liquid penetration for the temperature of \SI{700}{\kelvin} is slightly lower. On the other hand, no particular benefit is observed in the liquid lengths of \oct~and \dnbe, which are still underpredicted.
\subsubsection{Evaporation Model from Miller and Bellan}
\begin{figure}[]
	\centering
	\input{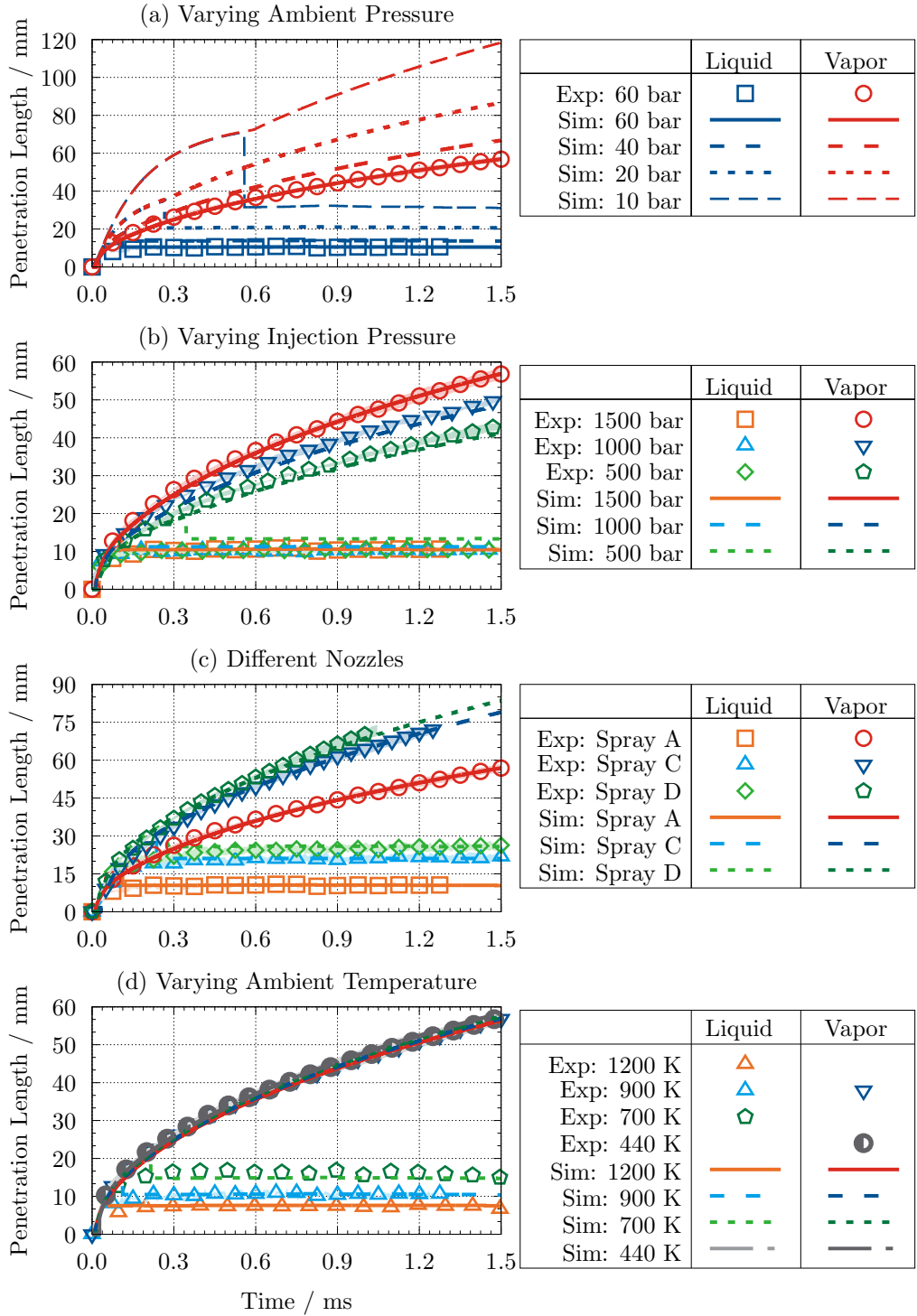}
	\caption{Results using the KH-RT breakup model, the transport equation for the vapor mass fraction, and the evaporation model of Miller and Bellan for cases 1-11.}
	\label{fig:miller_evap_ecn}
\end{figure}

The upgrade of the evaporation model to that of \cite{Miller1999} concludes the improvements in the monodisperse CAS model. For completeness, the results for all cases (1-15) are shown in Figures \ref{fig:miller_evap_ecn} and \ref{fig:miller_evap_fsc}. \myfig{fig:miller_evap_ecn}(d) shows the improvements in the liquid length for the ambient temperatures of \SI{700}{\kelvin} and \SI{900}{\kelvin}. The model prediction of the liquid penetration for \dnbe~and \oct~has improved significantly (see~\myfig{fig:miller_evap_fsc}) at both operating conditions.
\begin{figure}[!ht]
	\centering
	\input{./figures/NewCasFuel.tex}
	\caption{Results using the KH-RT breakup model, the transport equation for the vapor mass fraction, and the evaporation model of Miller and Bellan for cases 12-15.}
	\label{fig:miller_evap_fsc}
\end{figure}
\begin{figure}[!ht]
	\centering
	\input{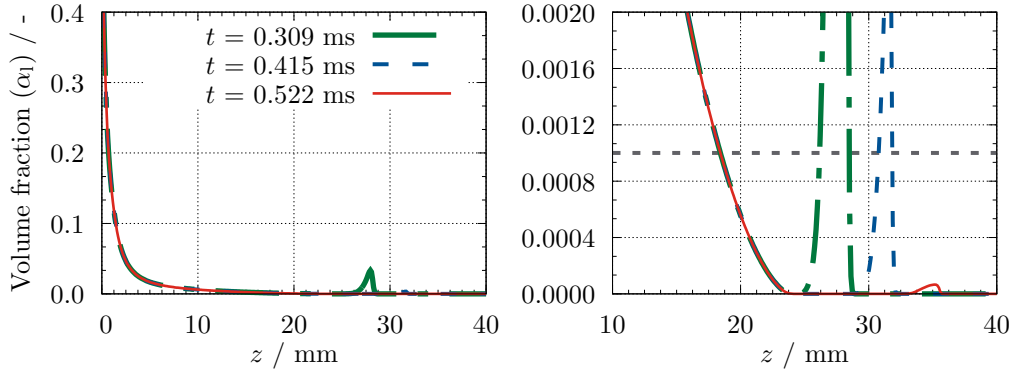}
	\caption{Liquid volume fraction at successive times (left), zoomed view (right) for case 12. Horizontal dashed gray line indicates the criterion of the liquid volume fraction ($\alphal$) = 0.1$\%$ for the liquid penetration length.}
	\label{fig:liq_vol_frac}
\end{figure}\par
The liquid penetration from the CAS model shows a jump (e.g., in~\myfig{fig:miller_evap_fsc}), which can be understood by observing the volume fraction of the liquid phase because the liquid length is determined as the farthest location from the nozzle exit where the liquid volume fraction falls below \SI{0.1}{\percent}. \myfig{fig:liq_vol_frac} shows the successive liquid volume fraction profiles along the axis of the spray for case 12 as an example. The localized bump at the tip of the spray is formed due to the initial transient ramp of the needle motion. As the initial velocity of the jet is lower than the steady-state velocity, breakup and evaporation of the large liquid mass are slow in the beginning. At the steady-state injection rate, the bump gradually flattens, and the liquid length jumps to lower values as soon as the bump falls below the threshold of \SI{0.1}{\percent}.
\begin{figure}[!ht]
	\centering
	\input{./figures/NewCasFlowVar.tex}
	\caption{Comparison of non-dimensional flow variables predicted with the CAS model and the URANS simulation at $t$ = \SI{1}{\milli\second} and the time-averaged LES for case 1.}
	\label{fig:miller_evap_flow_var}
\end{figure}
\subsection{Comparison with Simulation Data}
In addition to validation with the experimental data, the improved CAS model is compared in detail with the LES and URANS results. \myfig{fig:miller_evap_flow_var} shows the non-dimensional flow variables from the CAS model, the URANS simulation, and the LES along the non-dimensional axial coordinate (defined in ~\mytab{tab:init_bc}). The two-phase mixture density ($\rhostar$), mass fractions of the liquid ($\Ystarl$) and the ambient gas ($\Ystara$), and velocity of the liquid phase ($\ustarl$) and the gas phase ($\ustarg$) along the axial coordinate are shown in \myfig{fig:miller_evap_flow_var}(a). Due to the Lagrangian description, all variables are not directly comparable with those from the LES and URANS simulations.\par
The gas-phase mixture fraction (or vapor mass fraction in the inert case) is available in the ECN experimental data, the URANS results, and the LES results for comparison. Its cross-sectional average is compared with the mixture fraction from the CAS model (see \myfig{fig:miller_evap_flow_var}(b)). Until $t=$ \SI{1}{\milli\second}, the vapor in the URANS simulation has traveled to a shorter distance compared to the LES as well as the CAS model, which is consistent with the lower penetration length observed in the URANS simulation in \myfig{fig:org_cas_validation}. In the LES, the values are averaged in time during the steady-state phase from $t=$ \SIrange{1}{2}{\milli\second}. All of the simulation models are able to reasonably represent the mixture fraction away from the nozzle, although there are differences near the nozzle. The results of the CAS model are closer to the URANS results rather than the LES results, as expected given the averaged approach of the CAS model.\par
\myfig{fig:miller_evap_flow_var}(c) shows the mean droplet diameters along the axis of the spray from the LES, URANS, and CAS models. In the LES, the mean droplet diameter drops rapidly within a distance of $\approx$ $10\Dnoz$. In contrast, the mean droplet diameter in the URANS simulation decreases gradually over the liquid length. The mean droplet diameter in the CAS model shows trends similar to the LES but different from URANS, although a very similar KH-RT breakup model is used in all simulation approaches.\par
The mean droplet temperature in the CAS model follows the URANS result closely near the nozzle exit but deviates at farther distances, as shown in \myfig{fig:miller_evap_flow_var}(d). On the other hand, the droplets in the LES are heated much faster close to the nozzle, as the mean droplet diameter is much smaller compared to other simulation models. After $\approx$ $10\Dnoz$, the mean droplet temperature rises more gradually in the LES. The reasons for such differences can be attributed to different modeling of droplet dynamics in the CAS model. While droplets of different sizes can exist in the LES and URANS simulations, the CAS model assumes them to be monodisperse. Other factors include averaged modeling in CAS, which ignores the spatial distribution of droplets and hot ambient gas in the radial direction and the non-linear temperature dependency of the evaporation term. In the LES and URANS simulations, high temperatures exist in the outer regions and cold temperatures in the center. Droplets that move to the outer part evaporate quickly, while those in the center evaporate slowly. Such temperature distribution can affect the evaporation time, which may not be captured by the cross-sectional averaging.
\subsection{Droplet Size Distribution}
The CAS model was further extended in an attempt to obtain information on the droplet size distribution in the spray. Monodispersed droplet distribution was used for results in the previous sections by setting the PDF $\mathcal{P}(\dhat)$ to a delta function. As discussed before, the initial droplet size distribution at the nozzle exit must be known to use other PDFs. At high injection pressures and high ambient densities, the liquid fuel jet breaks into fine droplets very close to the nozzle, $\approx$ $10\Dnoz$, and therefore the secondary breakup processes may be negligible. For this reason, the KH-RT breakup model can be switched off, and only drag and evaporation govern the droplet dynamics, as shown by~\cite{Davidovic2017} in their LES results. For comparability with these results, the breakup model was switched off in the CAS model and the URANS simulation for the case discussed in this section. In addition, $\Cvap$ was set to unity in this case to reduce the evaporation rate.\par
\begin{figure}[!b]
	\centering
	\input{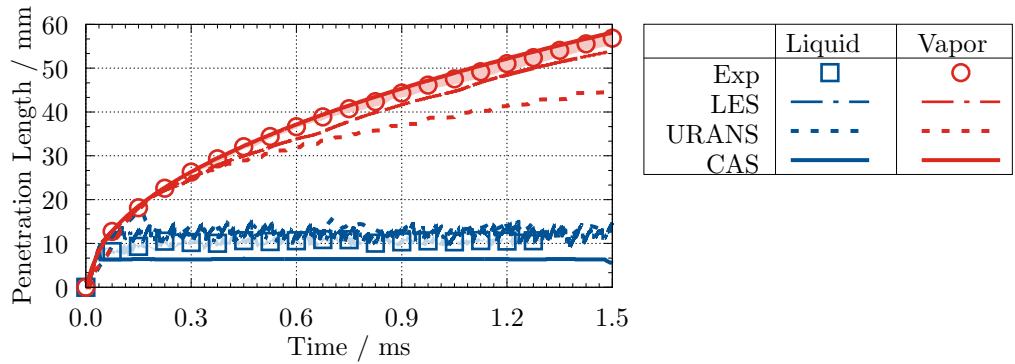}
	\caption{Comparison of the liquid and vapor penetration lengths for the LES, the URANS, and the CAS model with the Rosin-Rammler initial droplet size distribution (case 1).}
	\label{fig:dsd_cas_pen}
\end{figure}
\myfig{fig:dsd_cas_pen} shows the comparison of liquid and vapor penetration lengths for the LES, the URANS, and the CAS model with Rosin-Rammler PDF as the initial droplet size distribution, which is obtained from the DNS of primary breakup of Spray A in~\cite{Davidovic2017}. The results of lognormal and gamma PDFs are not shown here, as they do not significantly differ from the Rosin-Rammler PDF. The liquid lengths from the LES and URANS simulations agree well with the measurements, whereas the CAS model underpredicts the liquid length. The lower liquid length in the CAS model is caused by the overestimation of the evaporation rate. The vapor penetration lengths from the LES and the CAS model match closely with that from the experiment. On the other hand, the URANS simulation largely underpredicts the vapor penetration for the reasons discussed before.\par
\begin{figure}[!ht]
	\centering
	\input{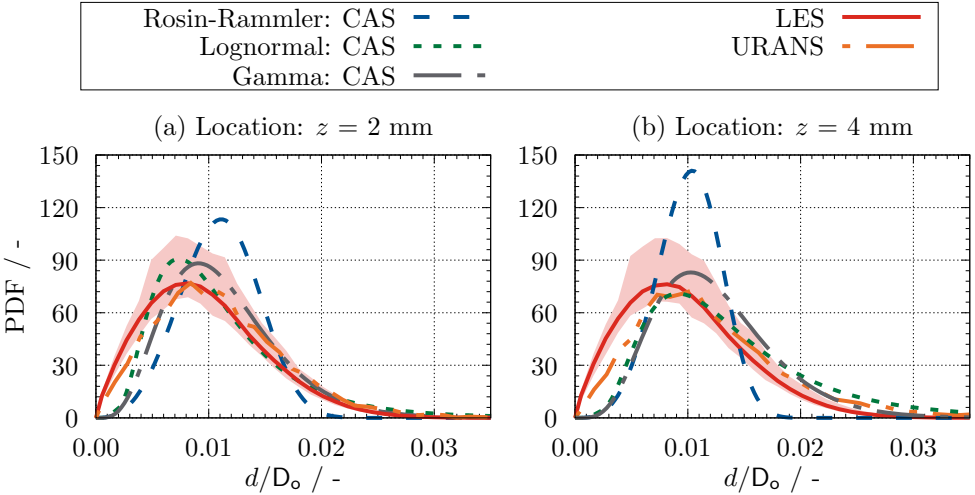}
	\caption{Comparison of the droplet size distributions for case 1 from the URANS simulation and the CAS model at $t$ = \SI{1}{\milli\second} on two locations away from the nozzle exit. The LES results are averaged in time ($t$ = \SIrange{1}{2}{\milli\second}), and the shaded region indicates the region of variation from minimum to maximum values.}
	\label{fig:dsdcompare}
\end{figure}

Within the liquid length, the droplet size distributions are compared in~\myfig{fig:dsdcompare} at two locations away from the nozzle. The shaded region indicates the variation within the steady-state period in the LES. Compared to the LES and URANS, the results of the CAS model with lognormal and gamma PDF represent the droplet size distribution at both locations reasonably. The Rosin-Rammler PDF is shifted to the right at 2\,mm and is narrower than others, which leads to the conclusion that it may not be suitable for tracking the droplet size distribution in the reduced-order CAS model. On the other hand, the lognormal and gamma PDFs represent the droplet size distributions reasonably, and therefore may be used in the proposed approach.\par
The modeling approach for the polydisperse droplet distribution is presented here for demonstration. Although the results are encouraging, further investigations are needed on the discrepancies regarding the liquid length, which will be part of future work.

\subsection{Computational Cost}\label{sec:comp_cost}
All simulations were run on machines using the Intel Broadwell processor architecture. The 3D inert URANS of Spray A was run on 24 cores for $\approx$ \SI{1}{\hour} for both delta and Rosin-Rammler initial droplet size distributions, resulting in a total of \SI{24}{CPUh}. The CAS model with the delta PDF for droplet size distribution was run on a single core for $\approx$ \SI{45}{\second} (= \SI{0.0125}{CPUh}) for the same case, which is a factor of 2000 compared to the URANS. For more complex droplet distributions, the computational cost of the CAS model increased to \SI{0.125}{CPUh} for the Rosin-Rammler PDF, \SI{0.8}{CPUh} for the gamma PDF, and \SI{1.2}{CPUh} for the lognormal PDF. The LES, on the other hand, required \SI{62}{\kilo CPUh} for the blob injection (delta PDF) and \SI{102}{\kilo CPUh} for the Rosin-Rammler initial droplet size distribution. Thus, the improved CAS model is faster by up to 3 orders of magnitude compared to the URANS and 4-6 orders of magnitude compared to the LES and provides reasonable estimates for liquid and vapor lengths, mixing, as well as droplet size distribution for a wide range of conditions and potential novel bio-hybrid fuels.

\section{Conclusions}\label{sec:conclude}
In this paper, the original CAS model was evaluated with respect to experiments, LES, and URANS results to determine its capabilities to predict macroscopic spray characteristics such as liquid and vapor penetration, mean droplet diameter, and mixing. The original model performed well for different ambient densities, injection pressures, and nozzles; however, it failed to predict the trends in the spray characteristics for variation in ambient temperatures and different fuel properties for the given injector. Also, the model in its original form did not track the vapor mass fraction, which plays an important role in the evaporation model and provides information on local mixing with the ambient gas. In this work, the model was improved by replacing the Reitz-Diwakar breakup model with the state-of-the-art Kelvin-Helmholtz Rayleigh-Taylor model. An additional equation is solved for the fuel-vapor transport required for the evaporation model, and the non-equilibrium evaporation model proposed by Miller and Bellan was implemented. In addition, a phenomenological nozzle flow model was implemented to predict the nozzle exit velocity. The improved CAS model was able to reproduce the trends in global spray characteristics for different ambient temperatures as well as novel bio-hybrid fuel candidates.\par
The model was further extended to consider polydisperse sprays. A transport equation was derived for the mean droplet diameter, in addition to the existing equation for the squared droplet diameter, and the droplet size distribution was constructed from the presumed PDFs, namely Rosin-Rammler, lognormal, gamma distributions, fitted using the two moments of the droplet diameter. This approach was able to track the droplet size distribution along the axial coordinate of the spray, and lognormal and gamma distributions were found to be more suitable rather than the commonly used Rosin-Rammler distribution.\par
The CAS model was developed to be a physics-based reduced-order model for various applications, which require parameter variation, sensitivity analysis, and real-time prediction, e.g., a rapid screening tool for a large number of novel bio-hybrid fuel candidates and model-based control of an engine. The 3D simulation methods are computationally expensive and hence not suitable for these applications. A posteriori computational cost analysis and comparison with 3D simulation methods showed that the improved CAS model with monodispersed droplets is faster by up to 3 orders of magnitude compared to the URANS and by up to 6 orders of magnitude compared to the LES. Modeling of the droplet size distribution increased the computational cost of the CAS model by a factor of 20-200 depending on the type of PDF used, with lognormal distribution as the most expensive and Rosin-Rammler as the least expensive distribution to compute.

\section*{Acknowledgements}
This work was funded by the Deutsche Forschungsgemeinschaft (DFG, German Research Foundation) under Germany's Excellence Strategy -- Exzellenzcluster 2186 "The Fuel Science Center" ID: 390919832. The authors gratefully acknowledge generous support of the European Research Council (ERC) under the European Union’s Horizon 2020 research and innovation program (grant agreement no. 695747). The authors are thankful to Convergent Science Inc. for providing licenses for CONVERGE. The authors gratefully acknowledge the computing time granted by the JARA Vergabegremium and provided on the JARA Partition part of the supercomputer CLAIX at RWTH Aachen University.

\appendix
\section{Original CAS Model}\label{sec:app_org_model}
\subsection{Governing Equations}
 The system of hyperbolic partial differential equations (\ac{PDE}s) for the original CAS model of~\cite{Wan1997} are written as
\begin{align}
		\Dg (\rhohat \Yhatg b^2) & = \wentg b  +  \wvap b^2 \label{eq:old_gas_cont} \\		
		\Dg (\rhohat \Yhatg \uhatg b^2) & = -\wdrag b^2 + \wvap \uhatl  b^2 \label{eq:old_gas_mom} \\
		\Dl (\rhohat \Yhatl b^2) & =  - \wvap b^2 \label{eq:old_liq_cont} \\
		\Dl (\rhohat \Yhatl \uhatl b^2) & = \wdrag b^2 - \wvap \uhatl  b^2 \label{eq:old_liq_mom} \\
		\Dl (\rhohat \Yhatl \dhat^2 b^2) & = -\wbre b^2 - \frac{5}{3}\wvap \dhat^2  b^2 \label{eq:old_droptrans} \\
		\Dl (\rhohat \Yhatl \Thatd b^2) & =   \wheat b^2 - \wvap  \Thatd b^2, \label{eq:old_liq_energy}
		%\label{eq:cas}
\end{align}
The two-phase mixture density is computed as 
\begin{align}
		\frac{1}{\rhohat} & = \frac{\Yhatl}{\rhol} + \frac{\Yhatg}{\rhog}.
		\label{eq:old_mix_den}
\end{align} 
Mass conservation requires $\Yhatl+\Yhatg = 1$. The entrainment and the drag models were described in sections~\mysec{sec:ent_model} and \mysec{sec:drag_model}, respectively. Since the vapor and ambient gas are considered to be perfectly mixed and vapor mass fraction is not tracked explicitly, the gas mixture is denoted by subscript '\ac{g}'. The droplet breakup and the evaporation models are described briefly in the next sections.
\subsection{Droplet Breakup Model}\label{sec:app_bre_model}
The~\cite{Reitz1986} (\ac{RD}) wave breakup model was used for the secondary breakup, assuming that most breakups were of stripping-type~\citep{Reitz1987}. The droplet breakup source term is computed as $\wbre = \Kbre \rhohat \Yhatl$, where the breakup coefficient, $\Kbre$, is modeled as
\begin{align}
		\Kbre  & = \frac{2\dhat (\dhat-\dstRD)}{\taubRD},
		\label{eq:old_kbre}
\end{align}
and the stable diameter is computed as
\begin{align}
		\dstRD & = \frac{\sigmal^2}{\rhog \ucuberel \mug},
		\label{eq:old_dst}
\end{align}
where $\sigma_l$ is the surface tension of the liquid. The stripping breakup time is computed as
\begin{align}
		\taubRD = \Cb \frac{\dhat}{\urel} \left(\frac{\rhol}{\rhog}\right)^{1/2},
		\label{eq:old_tb}
\end{align}
where $\Cb$ is a model constant with a suggested value of 10~\citep{Reitz1986}.
\subsection{Evaporation and Droplet Heating Model}\label{sec:app_evap_model}
\cite{Frossling1938} correlations are used to model evaporation from the external droplet surface. The evaporation source term is computed as
\begin{align}
		\wvap & = \frac{3 \Kvap \rhohat \Yhatl}{2\dhat^2},
		\label{eq:old_wvap}
\end{align} 
where
\begin{align}
		\Kvap &= 8 \frac{\rhog \Gvg}{\rhol}\;\mathrm{ln} (1+\BMd) \Shd.
		\label{eq:old_kvap}
\end{align} 
$\Gvg$ is the diffusion coefficient of fuel vapor (subscript '\ac{v}') in the surrounding gas mixture. The droplet mass transfer number is defined as  
\begin{align}
		\BMd & = \frac{\Ytvs-\Yiv}{1-\Ytvs},
		\label{eq:old_masstransferno}
\end{align}
where the vapor mass fraction on the droplet surface is modeled as
\begin{align}
		\Ytvs &  = \frac{\chieqvs}{\chieqvs +  \left(1 - \chieqvs\right) \mathsf{WR}},
		\label{eq:old_yvs}
\end{align}
where $\mathsf{WR} = W_{\mathrm{a}}/W_{\mathrm{f}}$ is the ratio of the molecular weight of the ambient gas excluding vapor (subscript '\ac{a}') to the molecular weight of fuel (subscript '\ac{f}'). The subscript '\ac{s}' denotes variables at the droplet surface. The equilibrium vapor mole fraction, $\chieqvs$, at the droplet surface is computed as
\begin{align}
	\chieqvs & =  \frac{\pv(\Thatd)}{\Pg},
	\label{eq:old_chieq}
\end{align}
where $\Pg$ is ambient gas pressure and $\pv(\Thatd)$ the vapor pressure at the droplet temperature. For the cases considered here, $\Yiv$ is set according to the one-third rule~\citep{Hubbard1975} to $\Yiv = (\Yhatv +2 \Ytvs)/3$, where $\Yhatv$ is chosen to be zero, assuming the far-field value of the vapor mass fraction to be negligible.
The droplet Sherwood number is computed as
\begin{align}
	\Shd & =  (2.0 + 0.6\;\Red^{1/2}\;\Scg^{1/3})\frac{\mathrm{ln}(1+\BMd)}{\BMd}
	 \label{eq:old_sherwoodno}
\end{align}
with the gas-phase Schmidt number, $\Scg$, defined as
\begin{align}
\Scg = \frac{\mug}{\rhog \Gvg}.
\end{align}
The droplet temperature is affected by continuous heating and evaporation. The respective source term is modeled as
$\wheat = \Kheat \rhohat \Yhatl$ with the temperature coefficient, $\Kheat$, given by
\begin{align}
		\Kheat  = \frac{6\Qd}{\rhol \dhat \Cl(\Thatd)} - \frac{3 \Kvap L(\Thatd)}{2 \dhat^2 \Cl(\Thatd)},
		\label{eq:old_kt}
\end{align}
where $L(\Thatd)$ and $\Cl(\Thatd)$ are the latent heat of vaporization and the heat capacity of the fuel at the droplet temperature $\Thatd$, respectively. The heat transfer between the droplet and ambient gas mixture is modeled as
\begin{align}
	\Qd = \frac{\lambdag(\Tref) (\Thatg - \Thatd)}{\dhat}\Nud,
	\label{eq:old_qd}
\end{align}
where $\lambdag(\Tref)$ is the thermal conductivity of the local gas-phase mixture and $\Thatg$ the ambient gas temperature.
The droplet Nusselt number is given by
\begin{align}
	\Nud = (2.0 + 0.6\;\Red^{1/2}\;\Prg^{1/3})\frac{\mathrm{ln}(1+\BMd)}{\BMd}, 
	\label{eq:old_nud}
\end{align}
and the gas-phase Prandtl number, $\Prg$, is defined as
\begin{align}
\Prg = \frac{\mug(\Tref) \Cpg(\Tref)}{\lambdag(\Tref)}.
\label{eq:old_prg}
\end{align}
$\Cpg(\Tref)$ is the specific heat capacity of the gas-phase mixture at constant pressure. The gas-phase mixture properties and correlations are computed at the reference temperature, which is obtained by the one-third rule~\citep{Hubbard1975}
\begin{align}
	\Tref = \frac{\Thatg+2\Thatd}{3}.
	\label{eq:old_tref}
\end{align}
Since the boiling model is not used, the droplet temperature is kept bounded by an upper limit of saturation temperature at the given pressure. The ambient gas temperature is held constant throughout the domain at all times. The liquid properties such as density ($\rhol$), surface tension ($\sigmal$), and viscosity ($\mul$) are evaluated at a constant fuel temperature at the nozzle exit ($\Tlo$) and remain constant throughout the liquid phase. The gas-phase mixture properties, such as density ($\rhog$), viscosity ($\mug$), thermal conductivity ($\lambdag$), and specific heat capacity at constant pressure ($\Cpg$) are approximated by the corresponding properties of ambient gas only.
\subsection{Comparison with LES and URANS}\label{sec:app_results}
\begin{figure}[!ht]
	\centering
	\input{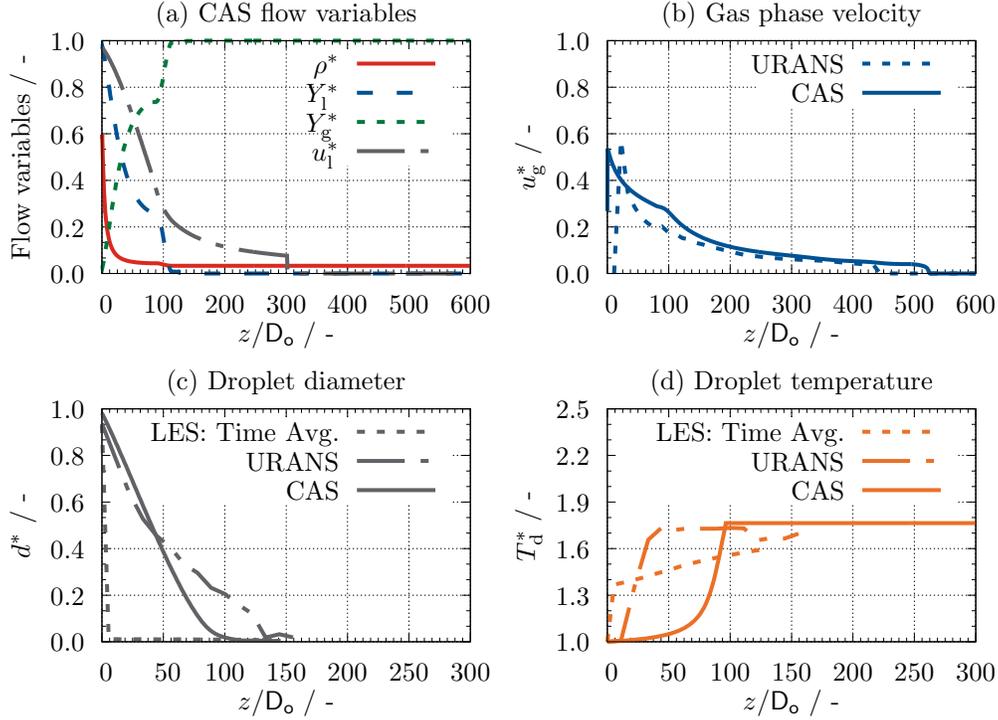}
	\caption{Comparison of non-dimensional flow variables predicted with the CAS model and the URANS simulation at $t$ = 1\,ms and the time-averaged LES.}
	\label{fig:org_cas_flow_var}
\end{figure}

\myfig{fig:org_cas_flow_var} shows the non-dimensional flow variables from the CAS model and the URANS simulation along the non-dimensional axial coordinate (defined in ~\mytab{tab:init_bc}). The two-phase mixture density ($\rhostar$), mass fractions of the liquid ($\Ystarl$) and the ambient gas ($\Ystarg$), and velocity of the liquid phase ($\ustarl$) along the axial coordinate are shown in \myfig{fig:org_cas_flow_var}(a). Due to the Lagrangian description, these variables are not directly comparable with those from the LES and URANS simulations.
\par
The gas-phase velocity is available in the URANS results for comparison, and its cross-sectional average is compared with the ambient gas velocity from the CAS model (see \myfig{fig:org_cas_flow_var}(b)). The LES results of the gas-phase velocity are not compared, as an ensemble averaging of several independent runs would be required, and these data are not available. High gas velocities are induced close to the nozzle in the CAS model, whereas velocities of similar magnitude occur in the URANS simulation at a farther location from the nozzle exit. The decay of the gas velocities in URANS is faster compared to that in the CAS model. These observations confirm the reasons discussed before for the underprediction of the vapor penetration in the URANS approach.\par
\myfig{fig:org_cas_flow_var}(c) shows the mean droplet diameters along the axis of the spray from the LES, URANS, and CAS models. The LES results are averaged in time during the steady-state phase. In the LES, the mean droplet diameter drops rapidly within a distance of $\approx$ $10\Dnoz$. In contrast, the mean droplet diameter in the URANS simulation decreases gradually over the liquid length. The CAS model shows trends similar to the URANS results, although the breakup models are quite different in the two simulation models, with combined KH-RT model in URANS and Reitz-Diwakar model in CAS.\par
The droplets in the LES and the URANS simulation heat up faster than those in the CAS model (see \myfig{fig:org_cas_flow_var}(d)). Also, the droplets in the CAS model survive up to much farther distances from the nozzle. These discrepancies can be attributed to the monodispersed droplet assumption at each axial location in the CAS model, whereas droplets of different sizes can exist in the LES and URANS simulations. Also, the non-linear temperature dependency of the evaporation term is a possible second reason.

\bibliography{references}

\begin{thebibliography}{91}
\expandafter\ifx\csname natexlab\endcsname\relax\def\natexlab#1{#1}\fi
\providecommand{\url}[1]{\texttt{#1}}
\providecommand{\href}[2]{#2}
\providecommand{\path}[1]{#1}
\providecommand{\DOIprefix}{doi:}
\providecommand{\ArXivprefix}{arXiv:}
\providecommand{\URLprefix}{URL: }
\providecommand{\Pubmedprefix}{pmid:}
\providecommand{\doi}[1]{\href{http://dx.doi.org/#1}{\path{#1}}}
\providecommand{\Pubmed}[1]{\href{pmid:#1}{\path{#1}}}
\providecommand{\bibinfo}[2]{#2}
\ifx\xfnm\relax \def\xfnm[#1]{\unskip,\space#1}\fi
%Type = Article
\bibitem[{Abraham(1997)}]{Abraham1997}
\bibinfo{author}{Abraham, J.}, \bibinfo{year}{1997}.
\newblock \bibinfo{title}{{What is adequate resolution in the numerical
  computations of transient jets?}}
\newblock \bibinfo{journal}{SAE Technical Paper 970051}
  \DOIprefix\doi{10.4271/970051}.
%Type = Article
\bibitem[{Agarwal and Trujillo(2020)}]{Agarwal2019}
\bibinfo{author}{Agarwal, A.}, \bibinfo{author}{Trujillo, M.F.},
  \bibinfo{year}{2020}.
\newblock \bibinfo{title}{{The effect of nozzle internal flow on spray
  atomization}}.
\newblock \bibinfo{journal}{International Journal of Engine Research}
  \bibinfo{volume}{21}, \bibinfo{pages}{55--72}.
\newblock \DOIprefix\doi{10.1177/1468087419875843}.
%Type = Article
\bibitem[{Albrecht et~al.(2007)Albrecht, Grondin, Berr and
  Solliec}]{Albrecht2007}
\bibinfo{author}{Albrecht, A.}, \bibinfo{author}{Grondin, O.},
  \bibinfo{author}{Berr, F.L.}, \bibinfo{author}{Solliec, G.L.},
  \bibinfo{year}{2007}.
\newblock \bibinfo{title}{{Towards a Stronger Simulation Support for Engine
  Control Design: a Methodological Point of View}}.
\newblock \bibinfo{journal}{Oil {\&} Gas Science and Technology – Revue d'IFP
  Energies nouvelles} \bibinfo{volume}{62}, \bibinfo{pages}{437--456}.
\newblock \DOIprefix\doi{10.2516/ogst:2007039}.
%Type = Article
\bibitem[{Beam and Warming(1976)}]{Beam1976}
\bibinfo{author}{Beam, R.M.}, \bibinfo{author}{Warming, R.F.},
  \bibinfo{year}{1976}.
\newblock \bibinfo{title}{{An implicit finite-difference algorithm for
  hyperbolic systems in conservation-law form}}.
\newblock \bibinfo{journal}{Journal of Computational Physics}
  \bibinfo{volume}{22}, \bibinfo{pages}{87--110}.
\newblock \DOIprefix\doi{10.1016/0021-9991(76)90110-8}.
%Type = Article
\bibitem[{B{\'{e}}ard et~al.(2000)B{\'{e}}ard, Duclos, Habchi, Bruneaux,
  Mokaddem and Baritaud}]{Beard2000}
\bibinfo{author}{B{\'{e}}ard, P.}, \bibinfo{author}{Duclos, J.M.},
  \bibinfo{author}{Habchi, C.}, \bibinfo{author}{Bruneaux, G.},
  \bibinfo{author}{Mokaddem, K.}, \bibinfo{author}{Baritaud, T.},
  \bibinfo{year}{2000}.
\newblock \bibinfo{title}{{Extension of lagrangian-eulerian spray modeling:
  Application to high pressure evaporating diesel sprays}}.
\newblock \bibinfo{journal}{SAE Technical Paper 2000-01-1893}
  \DOIprefix\doi{10.4271/2000-01-1893}.
%Type = Article
\bibitem[{Bengtsson et~al.(2007)Bengtsson, Strandh, Johansson, Tunest\r{a}l and
  Johansson}]{Bengtsson2007}
\bibinfo{author}{Bengtsson, J.}, \bibinfo{author}{Strandh, P.},
  \bibinfo{author}{Johansson, R.}, \bibinfo{author}{Tunest\r{a}l, P.},
  \bibinfo{author}{Johansson, B.}, \bibinfo{year}{2007}.
\newblock \bibinfo{title}{{Hybrid modelling of homogeneous charge compression
  ignition (HCCI) engine dynamics—a survey}}.
\newblock \bibinfo{journal}{International Journal of Control}
  \bibinfo{volume}{80}, \bibinfo{pages}{1814--1847}.
\newblock \DOIprefix\doi{10.1080/00207170701484869}.
%Type = Phdthesis
\bibitem[{Blanco(2016)}]{Blanco2016}
\bibinfo{author}{Blanco, A.}, \bibinfo{year}{2016}.
\newblock \bibinfo{title}{{Implementation and Development of an Eulerian Spray
  Model for CFD Simulations of Diesel Sprays}}.
\newblock Ph.D. thesis. Polytechnic University of Valencia.
%Type = Techreport
\bibitem[{Bravo and Kweon(2014)}]{Bravo2014}
\bibinfo{author}{Bravo, L.}, \bibinfo{author}{Kweon, C.B.},
  \bibinfo{year}{2014}.
\newblock \bibinfo{title}{{A Review on Liquid Spray Models for Diesel Engine
  Computational Analysis}}.
\newblock \bibinfo{type}{Technical Report}. Army Research Laboratory.
%Type = Misc
\bibitem[{{Co-Optima}(2020)}]{COOPTIMA2020}
\bibinfo{author}{{Co-Optima}}, \bibinfo{year}{2020}.
\newblock \bibinfo{title}{{Co-Optimization of Fuels and Engines}}.
\newblock \URLprefix
  \url{https://www.energy.gov/eere/bioenergy/co-optimization-fuels-engines}.
  \bibinfo{note}{{Accessed on 12.10.2020}}.
%Type = Book
\bibitem[{Crowe et~al.(2012)Crowe, Schwarzkopf, Sommerfeld and
  Tsuji}]{Crowe2012}
\bibinfo{author}{Crowe, C.T.}, \bibinfo{author}{Schwarzkopf, J.D.},
  \bibinfo{author}{Sommerfeld, M.}, \bibinfo{author}{Tsuji, Y.},
  \bibinfo{year}{2012}.
\newblock \bibinfo{title}{{Mulitphase Flows with Droplets and Particles}}.
\newblock \bibinfo{publisher}{Taylor \& Francis Group, LLC}.
%Type = Book
\bibitem[{Daubert and Danner(1989)}]{Daubert1989}
\bibinfo{author}{Daubert, T.E.}, \bibinfo{author}{Danner, R.P.},
  \bibinfo{year}{1989}.
\newblock \bibinfo{title}{{Physical and Thermodynamic Properties of Pure
  Compounds: Data Compilation}}.
\newblock \bibinfo{publisher}{New York : Hemisphere Pub. Corp}.
%Type = Article
\bibitem[{Davidovic et~al.(2017)Davidovic, Falkenstein, Bode, Cai, Kang,
  Hinrichs and Pitsch}]{Davidovic2017}
\bibinfo{author}{Davidovic, M.}, \bibinfo{author}{Falkenstein, T.},
  \bibinfo{author}{Bode, M.}, \bibinfo{author}{Cai, L.}, \bibinfo{author}{Kang,
  S.}, \bibinfo{author}{Hinrichs, J.}, \bibinfo{author}{Pitsch, H.},
  \bibinfo{year}{2017}.
\newblock \bibinfo{title}{{LES of n -Dodecane Spray Combustion Using a Multiple
  Representative Interactive Flamelets Model}}.
\newblock \bibinfo{journal}{Oil and Gas Science and Technology - Rev. IFP
  Energies nouvelles} \bibinfo{volume}{72}.
\newblock \DOIprefix\doi{10.2516/ogst/2017019}.
%Type = Article
\bibitem[{Dent(1971)}]{Dent1971}
\bibinfo{author}{Dent, J.C.}, \bibinfo{year}{1971}.
\newblock \bibinfo{title}{{A basis for the comparison of various experimental
  methods for studying spray penetration}}.
\newblock \bibinfo{journal}{SAE Technical Paper 710571}
  \DOIprefix\doi{10.4271/710571}.
%Type = Article
\bibitem[{Desantes et~al.(2020)Desantes, García-Oliver, Pastor, Olmeda, Pandal
  and Naud}]{Desantes2020}
\bibinfo{author}{Desantes, J.}, \bibinfo{author}{García-Oliver, J.},
  \bibinfo{author}{Pastor, J.}, \bibinfo{author}{Olmeda, I.},
  \bibinfo{author}{Pandal, A.}, \bibinfo{author}{Naud, B.},
  \bibinfo{year}{2020}.
\newblock \bibinfo{title}{{LES Eulerian diffuse-interface modeling of fuel
  dense sprays near- and far-field}}.
\newblock \bibinfo{journal}{International Journal of Multiphase Flow}
  \bibinfo{volume}{127}, \bibinfo{pages}{103272}.
\newblock \DOIprefix\doi{10.1016/j.ijmultiphaseflow.2020.103272}.
%Type = Article
\bibitem[{Desantes et~al.(2009)Desantes, Pastor, Garc{\'{i}}a-Oliver and
  Pastor}]{Desantes2009}
\bibinfo{author}{Desantes, J.}, \bibinfo{author}{Pastor, J.},
  \bibinfo{author}{Garc{\'{i}}a-Oliver, J.}, \bibinfo{author}{Pastor, J.},
  \bibinfo{year}{2009}.
\newblock \bibinfo{title}{{A 1D model for the description of mixing-controlled
  reacting diesel sprays}}.
\newblock \bibinfo{journal}{Combustion and Flame} \bibinfo{volume}{156},
  \bibinfo{pages}{234--249}.
\newblock \DOIprefix\doi{10.1016/j.combustflame.2008.10.008}.
%Type = Article
\bibitem[{Desantes et~al.(2006)Desantes, Payri, Salvador and
  Gil}]{Desantes2006}
\bibinfo{author}{Desantes, J.M.}, \bibinfo{author}{Payri, R.},
  \bibinfo{author}{Salvador, F.J.}, \bibinfo{author}{Gil, A.},
  \bibinfo{year}{2006}.
\newblock \bibinfo{title}{{Development and validation of a theoretical model
  for diesel spray penetration}}.
\newblock \bibinfo{journal}{Fuel} \bibinfo{volume}{22},
  \bibinfo{pages}{87--110}.
\newblock \DOIprefix\doi{10.1016/j.fuel.2005.10.023}.
%Type = Article
\bibitem[{Desjardins et~al.(2008)Desjardins, Blanquart, Balarac and
  Pitsch}]{Desjardins2008}
\bibinfo{author}{Desjardins, O.}, \bibinfo{author}{Blanquart, G.},
  \bibinfo{author}{Balarac, G.}, \bibinfo{author}{Pitsch, H.},
  \bibinfo{year}{2008}.
\newblock \bibinfo{title}{{High order conservative finite difference scheme for
  variable density low Mach number turbulent flows}}.
\newblock \bibinfo{journal}{Journal of Computational Physics}
  \bibinfo{volume}{227}, \bibinfo{pages}{7125--7159}.
\newblock \DOIprefix\doi{10.1016/j.jcp.2008.03.027}.
%Type = Misc
\bibitem[{ECN(2020)}]{ECN2020}
\bibinfo{author}{ECN}, \bibinfo{year}{2020}.
\newblock \bibinfo{title}{{Engine Combustion Network (ECN)}}.
\newblock \URLprefix \url{https://ecn.sandia.gov/}. \bibinfo{note}{{Accessed on
  12.10.2020}}.
%Type = Misc
\bibitem[{{EU Comission}(2018)}]{EuropeanCommission2018}
\bibinfo{author}{{EU Comission}}, \bibinfo{year}{2018}.
\newblock \bibinfo{title}{{Communication from The Commission to The European
  Parliament, The European Council, The Council, The European Economic And
  Social Committee, The Committee of The Regions and The European Investment
  Bank A Clean Planet for all A European strategic long-term vision for a
  prosperous, modern, competitive and climate neutral economy}}.
\newblock \URLprefix
  \url{https://eur-lex.europa.eu/legal-content/EN/TXT/PDF/?uri=CELEX:52018DC0773\&from=EN}.
%Type = Article
\bibitem[{Feng et~al.(2019)Feng, Tang, Yin, Zhang and Huang}]{Feng2019}
\bibinfo{author}{Feng, Z.}, \bibinfo{author}{Tang, C.}, \bibinfo{author}{Yin,
  Y.}, \bibinfo{author}{Zhang, P.}, \bibinfo{author}{Huang, Z.},
  \bibinfo{year}{2019}.
\newblock \bibinfo{title}{{Time-resolved droplet size and velocity
  distributions in a dilute region of a high-pressure pulsed diesel spray}}.
\newblock \bibinfo{journal}{International Journal of Heat and Mass Transfer}
  \bibinfo{volume}{133}, \bibinfo{pages}{745--755}.
\newblock \DOIprefix\doi{10.1016/j.ijheatmasstransfer.2018.12.147}.
%Type = Article
\bibitem[{Fr{\"{o}}ssling(1938)}]{Frossling1938}
\bibinfo{author}{Fr{\"{o}}ssling, N.}, \bibinfo{year}{1938}.
\newblock \bibinfo{title}{{Uber die verdunstung fallender tropfen}}.
\newblock \bibinfo{journal}{Geophysics} \bibinfo{volume}{12},
  \bibinfo{pages}{170--216}.
%Type = Misc
\bibitem[{FSC(2020)}]{FSC2020}
\bibinfo{author}{FSC}, \bibinfo{year}{2020}.
\newblock \bibinfo{title}{{The Fuel Science Center (FSC)}}.
\newblock \URLprefix \url{http://www.fuelcenter.rwth-aachen.de/}.
  \bibinfo{note}{{Accessed on 12.10.2020}}.
%Type = Misc
\bibitem[{{Future Energy Systems}(2020)}]{FES2020}
\bibinfo{author}{{Future Energy Systems}}, \bibinfo{year}{2020}.
\newblock \bibinfo{title}{{Future Energy Systems}}.
\newblock \URLprefix
  \url{https://www.futureenergysystems.ca/research/sustainability/biomass}.
  \bibinfo{note}{{Accessed 12.10.2020}}.
%Type = Article
\bibitem[{Garcia-Oliver et~al.(2013)Garcia-Oliver, Pastor, Pandal, Trask,
  Baldwin and Schmidt}]{Garcia2013}
\bibinfo{author}{Garcia-Oliver, J.M.}, \bibinfo{author}{Pastor, J.M.},
  \bibinfo{author}{Pandal, A.}, \bibinfo{author}{Trask, N.},
  \bibinfo{author}{Baldwin, E.}, \bibinfo{author}{Schmidt, D.P.},
  \bibinfo{year}{2013}.
\newblock \bibinfo{title}{{Diesel Spray CFD Simulations based on the $\Sigma$-Y
  Eulerian Atomization}}.
\newblock \bibinfo{journal}{Atomization and Sprays} \bibinfo{volume}{23},
  \bibinfo{pages}{71--95}.
\newblock \DOIprefix\doi{10.1615/AtomizSpr.2013007198}.
%Type = Article
\bibitem[{Gorokhovski and Herrmann(2008)}]{Gorokhovski2008}
\bibinfo{author}{Gorokhovski, M.}, \bibinfo{author}{Herrmann, M.},
  \bibinfo{year}{2008}.
\newblock \bibinfo{title}{{Modeling Primary Atomization}}.
\newblock \bibinfo{journal}{Annual Review of Fluid Mechanics}
  \bibinfo{volume}{40}, \bibinfo{pages}{343--366}.
\newblock \DOIprefix\doi{10.1146/annurev.fluid.40.111406.102200}.
%Type = Article
\bibitem[{Han et~al.(2002)Han, Lu, Xie, Lai and Henein}]{Han2002}
\bibinfo{author}{Han, J.S.}, \bibinfo{author}{Lu, P.H.}, \bibinfo{author}{Xie,
  X.B.}, \bibinfo{author}{Lai, M.C.}, \bibinfo{author}{Henein, N.A.},
  \bibinfo{year}{2002}.
\newblock \bibinfo{title}{{Investigation of Diesel Spray Primary Break-up and
  Development for Different Nozzle Geometries}}.
\newblock \bibinfo{journal}{SAE Technical Paper 2002-01-2775}
  \DOIprefix\doi{10.4271/2002-01-2775}.
%Type = Inproceedings
\bibitem[{Hasse and Peters(2002)}]{Hasse2002}
\bibinfo{author}{Hasse, C.}, \bibinfo{author}{Peters, N.},
  \bibinfo{year}{2002}.
\newblock \bibinfo{title}{{Eulerian spray modeling of diesel injection in a
  highpressure/high temperature chamber}}, in: \bibinfo{booktitle}{11th
  International Multidimensional Engine Modeling User's Group Meeting},
  \bibinfo{publisher}{University of WisconsinMadison, Engine Research Center},
  \bibinfo{address}{Detroit}.
%Type = Inproceedings
\bibitem[{Hasse et~al.(2003)Hasse, Vogel and Peters}]{Hasse2003}
\bibinfo{author}{Hasse, C.}, \bibinfo{author}{Vogel, S.},
  \bibinfo{author}{Peters, N.}, \bibinfo{year}{2003}.
\newblock \bibinfo{title}{{Modeling of a DaimlerChrysler Truck Engine using an
  Eulerian Spray Model}}, in: \bibinfo{booktitle}{13th International
  Multidimensional Engine Modeling User's Group Meeting},
  \bibinfo{publisher}{Cray Inc.}
%Type = Article
\bibitem[{Hay and Jones(1972)}]{Hay1972}
\bibinfo{author}{Hay, N.}, \bibinfo{author}{Jones, P.L.}, \bibinfo{year}{1972}.
\newblock \bibinfo{title}{{Comparison of the various correlations for spray
  penetration}}.
\newblock \bibinfo{journal}{SAE Technical Paper 720776}
  \DOIprefix\doi{10.4271/720776}.
%Type = Article
\bibitem[{Helmers et~al.(2020)Helmers, Dietz and Weiss}]{Helmers2020}
\bibinfo{author}{Helmers, E.}, \bibinfo{author}{Dietz, J.},
  \bibinfo{author}{Weiss, M.}, \bibinfo{year}{2020}.
\newblock \bibinfo{title}{{Sensitivity analysis in the life-cycle assessment of
  electric vs. combustion engine cars under approximate real-world
  conditions}}.
\newblock \bibinfo{journal}{Sustainability (Switzerland)} \bibinfo{volume}{12},
  \bibinfo{pages}{1241}.
\newblock \DOIprefix\doi{10.3390/su12031241}.
%Type = Article
\bibitem[{Hillion et~al.(2009)Hillion, Buhlbuck, Chauvin and
  Petit}]{Hillion2009}
\bibinfo{author}{Hillion, M.}, \bibinfo{author}{Buhlbuck, H.},
  \bibinfo{author}{Chauvin, J.}, \bibinfo{author}{Petit, N.},
  \bibinfo{year}{2009}.
\newblock \bibinfo{title}{{Combustion Control of Diesel Engines Using Injection
  Timing}}.
\newblock \bibinfo{journal}{SAE Technical Paper 2009-01-0367}
  \DOIprefix\doi{10.4271/2009-01-0367}.
%Type = Article
\bibitem[{Hiroyasu and Arai(1980)}]{Hiroyasu1980}
\bibinfo{author}{Hiroyasu, H.}, \bibinfo{author}{Arai, M.},
  \bibinfo{year}{1980}.
\newblock \bibinfo{title}{{Fuel spray penetration and spray angle of diesel
  engines}}.
\newblock \bibinfo{journal}{Trans. JSAE} \bibinfo{volume}{21},
  \bibinfo{pages}{5--11}.
%Type = Article
\bibitem[{Hiroyasu and Kadota(1974)}]{Hiroyasu1974}
\bibinfo{author}{Hiroyasu, H.}, \bibinfo{author}{Kadota, T.},
  \bibinfo{year}{1974}.
\newblock \bibinfo{title}{{Fuel droplet size distribution in diesel combustion
  chamber}}.
\newblock \bibinfo{journal}{SAE Technical Paper 740715}
  \DOIprefix\doi{10.4271/740715}.
%Type = Article
\bibitem[{Hubbard et~al.(1975)Hubbard, Denny and Mills}]{Hubbard1975}
\bibinfo{author}{Hubbard, G.}, \bibinfo{author}{Denny, V.},
  \bibinfo{author}{Mills, A.}, \bibinfo{year}{1975}.
\newblock \bibinfo{title}{{Droplet evaporation: Effects of transients and
  variable properties}}.
\newblock \bibinfo{journal}{International Journal of Heat and Mass Transfer}
  \bibinfo{volume}{18}, \bibinfo{pages}{1003--1008}.
\newblock \DOIprefix\doi{10.1016/0017-9310(75)90217-3}.
%Type = Incollection
\bibitem[{{Intergovernmental Panel on Climate Change (IPCC)}(2014)}]{IPCC2014}
\bibinfo{author}{{Intergovernmental Panel on Climate Change (IPCC)}},
  \bibinfo{year}{2014}.
\newblock \bibinfo{title}{{Transport}}, in: \bibinfo{editor}{Edenhofer, O.},
  \bibinfo{editor}{Pichs-Madruga, R.}, \bibinfo{editor}{Sokona, Y.},
  \bibinfo{editor}{Farahani, E.}, \bibinfo{editor}{Kadner, S.},
  \bibinfo{editor}{Seyboth, K.}, \bibinfo{editor}{Adler, A.},
  \bibinfo{editor}{Baum, I.}, \bibinfo{editor}{Brunner, S.},
  \bibinfo{editor}{Eickemeier, P.}, \bibinfo{editor}{Kriemann, B.},
  \bibinfo{editor}{Savolainen, J.}, \bibinfo{editor}{Schl{\"{o}}mer, S.},
  \bibinfo{editor}{von Stechow, C.}, \bibinfo{editor}{Zwickel, T.},
  \bibinfo{editor}{Minx, J.} (Eds.), \bibinfo{booktitle}{Climate Change 2014
  Mitigation of Climate Change}. \bibinfo{publisher}{Cambridge University
  Press}, \bibinfo{address}{Cambridge}. chapter~\bibinfo{chapter}{8}, pp.
  \bibinfo{pages}{599--670}.
\newblock \DOIprefix\doi{10.1017/CBO9781107415416.014}.
%Type = Article
\bibitem[{Jiang and Shu(1996)}]{Jiang1996}
\bibinfo{author}{Jiang, G.}, \bibinfo{author}{Shu, C.}, \bibinfo{year}{1996}.
\newblock \bibinfo{title}{{Efficient implementation of weighted ENO schemes}}.
\newblock \bibinfo{journal}{Journal of computational physics}
  \bibinfo{volume}{228}, \bibinfo{pages}{202--228}.
\newblock \DOIprefix\doi{10.1006/jcph.1996.0130}.
%Type = Article
\bibitem[{Kahila et~al.(2018)Kahila, Wehrfritz, Kaario, {Ghaderi Masouleh},
  Maes, Somers and Vuorinen}]{Kahila2018}
\bibinfo{author}{Kahila, H.}, \bibinfo{author}{Wehrfritz, A.},
  \bibinfo{author}{Kaario, O.}, \bibinfo{author}{{Ghaderi Masouleh}, M.},
  \bibinfo{author}{Maes, N.}, \bibinfo{author}{Somers, B.},
  \bibinfo{author}{Vuorinen, V.}, \bibinfo{year}{2018}.
\newblock \bibinfo{title}{{Large-eddy simulation on the influence of injection
  pressure in reacting Spray A}}.
\newblock \bibinfo{journal}{Combustion and Flame} \bibinfo{volume}{191},
  \bibinfo{pages}{142--159}.
\newblock \DOIprefix\doi{10.1016/j.combustflame.2018.01.004}.
%Type = Inproceedings
\bibitem[{Korkmaz et~al.(2018a)Korkmaz, Golc, Jochim, Beeckmann and
  Pitsch}]{Korkmaz2018}
\bibinfo{author}{Korkmaz, M.}, \bibinfo{author}{Golc, D.},
  \bibinfo{author}{Jochim, B.}, \bibinfo{author}{Beeckmann, J.},
  \bibinfo{author}{Pitsch, H.}, \bibinfo{year}{2018}a.
\newblock \bibinfo{title}{{Development of a Fully Flexible Injection Strategy
  for Model-Based Combustion Control of PCCI Diesel Engine}}, in:
  \bibinfo{booktitle}{Symposium for Combustion Control},
  \bibinfo{address}{Aachen, Germany}.
%Type = Article
\bibitem[{Korkmaz et~al.(2018b)Korkmaz, Zweigel, Jochim, Beeckmann, Abel and
  Pitsch}]{Korkmaz2018a}
\bibinfo{author}{Korkmaz, M.}, \bibinfo{author}{Zweigel, R.},
  \bibinfo{author}{Jochim, B.}, \bibinfo{author}{Beeckmann, J.},
  \bibinfo{author}{Abel, D.}, \bibinfo{author}{Pitsch, H.},
  \bibinfo{year}{2018}b.
\newblock \bibinfo{title}{{Triple-injection strategy for modelbased control of
  premixed charge compression ignition diesel engine combustion}}.
\newblock \bibinfo{journal}{International Journal of Engine Research}
  \bibinfo{volume}{19}, \bibinfo{pages}{230--240}.
\newblock \DOIprefix\doi{10.1177/1468087417730485}.
%Type = Article
\bibitem[{Lax(1954)}]{Lax1954}
\bibinfo{author}{Lax, P.D.}, \bibinfo{year}{1954}.
\newblock \bibinfo{title}{{Weak solutions of nonlinear hyperbolic equations and
  their numerical computation}}.
\newblock \bibinfo{journal}{Communications on Pure and Applied Mathematics}
  \bibinfo{volume}{7}, \bibinfo{pages}{159--193}.
\newblock \DOIprefix\doi{10.1002/cpa.3160070112}.
%Type = Inproceedings
\bibitem[{MacCormack(1969)}]{MacCormack1969}
\bibinfo{author}{MacCormack, R.}, \bibinfo{year}{1969}.
\newblock \bibinfo{title}{{The effect of viscosity in hypervelocity impact
  cratering}}, in: \bibinfo{booktitle}{4th Aerodynamic Testing Conference},
  \bibinfo{publisher}{American Institute of Aeronautics and Astronautics},
  \bibinfo{address}{Reston, Virigina}.
\newblock \DOIprefix\doi{10.2514/6.1969-354}.
%Type = Article
\bibitem[{Matheis and Hickel(2018)}]{Matheis2018}
\bibinfo{author}{Matheis, J.}, \bibinfo{author}{Hickel, S.},
  \bibinfo{year}{2018}.
\newblock \bibinfo{title}{{Multi-component vapor-liquid equilibrium model for
  LES of high-pressure fuel injection and application to ECN Spray A}}.
\newblock \bibinfo{journal}{International Journal of Multiphase Flow}
  \bibinfo{volume}{99}, \bibinfo{pages}{294 --311}.
\newblock \DOIprefix\doi{10.1016/j.ijmultiphaseflow.2017.11.001}.
%Type = Article
\bibitem[{Miller and Bellan(1999)}]{Miller1999}
\bibinfo{author}{Miller, R.S.}, \bibinfo{author}{Bellan, J.},
  \bibinfo{year}{1999}.
\newblock \bibinfo{title}{{Direct numerical simulation of a confined
  three-dimensional gas mixing layer with one evaporating
  hydrocarbon-droplet-laden stream}}.
\newblock \bibinfo{journal}{Journal of Fluid Mechanics} \bibinfo{volume}{384},
  \bibinfo{pages}{293--338}.
\newblock \DOIprefix\doi{10.1017/S0022112098004042}.
%Type = Article
\bibitem[{Mittal et~al.(2014)Mittal, Kang, Doran, Cook and Pitsch}]{Mittal2014}
\bibinfo{author}{Mittal, V.}, \bibinfo{author}{Kang, S.},
  \bibinfo{author}{Doran, E.}, \bibinfo{author}{Cook, D.},
  \bibinfo{author}{Pitsch, H.}, \bibinfo{year}{2014}.
\newblock \bibinfo{title}{{LES of Gas Exchange in IC Engines}}.
\newblock \bibinfo{journal}{Oil {\&} Gas Science and Technology – Revue d'IFP
  Energies nouvelles} \bibinfo{volume}{69}, \bibinfo{pages}{29--40}.
\newblock \DOIprefix\doi{10.2516/ogst/2013122}.
%Type = Article
\bibitem[{Mugele and Evans(1951)}]{Mugele1951}
\bibinfo{author}{Mugele, R.A.}, \bibinfo{author}{Evans, H.D.},
  \bibinfo{year}{1951}.
\newblock \bibinfo{title}{{Droplet Size Distribution in Sprays}}.
\newblock \bibinfo{journal}{Industrial {\&} Engineering Chemistry}
  \bibinfo{volume}{43}, \bibinfo{pages}{1317--1324}.
\newblock \DOIprefix\doi{10.1021/ie50498a023}.
%Type = Article
\bibitem[{Musculus and Kattke(2009)}]{Musculus2009}
\bibinfo{author}{Musculus, M.P.}, \bibinfo{author}{Kattke, K.},
  \bibinfo{year}{2009}.
\newblock \bibinfo{title}{{Entrainment waves in diesel jets}}.
\newblock \bibinfo{journal}{SAE Technical Paper 2009-01-1355}
  \DOIprefix\doi{10.4271/2009-01-1355}.
%Type = Article
\bibitem[{Naber and Siebers(1996)}]{Naber1996}
\bibinfo{author}{Naber, J.}, \bibinfo{author}{Siebers, D.L.},
  \bibinfo{year}{1996}.
\newblock \bibinfo{title}{{Effects of Gas Density and Vaporization on
  Penetration and Dispersion of Diesel Sprays}}.
\newblock \bibinfo{journal}{SAE Technical Paper 960034}
  \DOIprefix\doi{10.4271/960034}.
%Type = Inproceedings
\bibitem[{Naber et~al.(1995)Naber, Siebers and Hencken}]{Naber1995}
\bibinfo{author}{Naber, J.D.}, \bibinfo{author}{Siebers, D.L.},
  \bibinfo{author}{Hencken, K.R.}, \bibinfo{year}{1995}.
\newblock \bibinfo{title}{{The Effects of High Ambient Density on Diesel Spray
  Penetration and Spread}}, in: \bibinfo{booktitle}{Joint Technical Meeting
  Central and Western States (USA) Sections of the International Combustion
  Institute}, \bibinfo{address}{San Antonio}. pp. \bibinfo{pages}{761--766}.
%Type = Article
\bibitem[{Nukiyama and Tanasawa(1939)}]{Nukiyama1939}
\bibinfo{author}{Nukiyama, S.}, \bibinfo{author}{Tanasawa, Y.},
  \bibinfo{year}{1939}.
\newblock \bibinfo{title}{{An Experiment on the Atomization of Liquid. : 4th
  Report, The Effect of the Properties of Liquid on the Size of Drops.}}
\newblock \bibinfo{journal}{Transactions of the Japan Society of Mechanical
  Engineers} \bibinfo{volume}{5}, \bibinfo{pages}{136--143}.
\newblock \DOIprefix\doi{10.1299/kikai1938.5.136}.
%Type = Article
\bibitem[{Palmer et~al.(2015)Palmer, Ramesh, Kirsch, Reddemann and
  Kneer}]{Palmer2015}
\bibinfo{author}{Palmer, J.}, \bibinfo{author}{Ramesh, M.},
  \bibinfo{author}{Kirsch, V.}, \bibinfo{author}{Reddemann, M.},
  \bibinfo{author}{Kneer, R.}, \bibinfo{year}{2015}.
\newblock \bibinfo{title}{{Spray Analysis of C8H18O Fuel Blends Using
  High-Speed Schlieren Imaging and Mie Scattering}}.
\newblock \bibinfo{journal}{SAE Technical Paper 2015-24-2478}
  \DOIprefix\doi{10.4271/2015-24-2478}.
%Type = Article
\bibitem[{Pastor et~al.(2015)Pastor, Garc{\'{i}}a-Oliver, Pastor and
  Vera-Tudela}]{Pastor2015}
\bibinfo{author}{Pastor, J.V.}, \bibinfo{author}{Garc{\'{i}}a-Oliver, J.M.},
  \bibinfo{author}{Pastor, J.M.}, \bibinfo{author}{Vera-Tudela, W.},
  \bibinfo{year}{2015}.
\newblock \bibinfo{title}{{One-dimensional diesel spray modeling of
  multicomponent fuels}}.
\newblock \bibinfo{journal}{Atomization and Sprays} \bibinfo{volume}{25},
  \bibinfo{pages}{485--517}.
\newblock \DOIprefix\doi{10.1615/AtomizSpr.2014010370}.
%Type = Article
\bibitem[{Pastor et~al.(2008)Pastor, {Javier Lopez}, Garcia and
  Pastor}]{Pastor2008}
\bibinfo{author}{Pastor, J.V.}, \bibinfo{author}{{Javier Lopez}, J.},
  \bibinfo{author}{Garcia, J.}, \bibinfo{author}{Pastor, J.M.},
  \bibinfo{year}{2008}.
\newblock \bibinfo{title}{{A 1D model for the description of mixing-controlled
  inert diesel sprays}}.
\newblock \bibinfo{journal}{Fuel} \bibinfo{volume}{87},
  \bibinfo{pages}{2871--2885}.
\newblock \DOIprefix\doi{10.1016/j.fuel.2008.04.017}.
%Type = Article
\bibitem[{Patterson and Reitz(1998)}]{Patterson1998}
\bibinfo{author}{Patterson, M.A.}, \bibinfo{author}{Reitz, R.D.},
  \bibinfo{year}{1998}.
\newblock \bibinfo{title}{{Modeling the Effects of Fuel Spray Characteristics
  on Diesel Engine Combustion and Emission}}.
\newblock \bibinfo{journal}{SAE Technical Paper 980131}
  \DOIprefix\doi{10.4271/980131}.
%Type = Misc
\bibitem[{Payri et~al.(2020)Payri, Salvador, Gimeno and Bracho}]{VIR2020}
\bibinfo{author}{Payri, R.}, \bibinfo{author}{Salvador, F.},
  \bibinfo{author}{Gimeno, J.}, \bibinfo{author}{Bracho, G.},
  \bibinfo{year}{2020}.
\newblock \bibinfo{title}{{Virtual Injection Rate Generator}}.
\newblock \URLprefix \url{https://www.cmt.upv.es/ECN03.aspx}.
  \bibinfo{note}{{Accessed on 12.10.2020}}.
%Type = Article
\bibitem[{Pickett et~al.(2010)Pickett, Genzale, Bruneaux, Malbec, Hermant,
  Christiansen and Schramm}]{Pickett2010}
\bibinfo{author}{Pickett, L.M.}, \bibinfo{author}{Genzale, C.L.},
  \bibinfo{author}{Bruneaux, G.}, \bibinfo{author}{Malbec, L.M.},
  \bibinfo{author}{Hermant, L.}, \bibinfo{author}{Christiansen, C.},
  \bibinfo{author}{Schramm, J.}, \bibinfo{year}{2010}.
\newblock \bibinfo{title}{{Comparison of Diesel Spray Combustion in Different
  High-Temperature, High-Pressure Facilities}}.
\newblock \bibinfo{journal}{SAE International Journal of Engines}
  \bibinfo{volume}{3}, \bibinfo{pages}{156--181}.
\newblock \DOIprefix\doi{10.4271/2010-01-2106}.
%Type = Article
\bibitem[{Pickett et~al.(2011)Pickett, Manin, Genzale, Siebers, Musculus and
  Idicheria}]{Pickett2011}
\bibinfo{author}{Pickett, L.M.}, \bibinfo{author}{Manin, J.},
  \bibinfo{author}{Genzale, C.L.}, \bibinfo{author}{Siebers, D.L.},
  \bibinfo{author}{Musculus, M.P.}, \bibinfo{author}{Idicheria, C.A.},
  \bibinfo{year}{2011}.
\newblock \bibinfo{title}{{Relationship Between Diesel Fuel Spray Vapor
  Penetration/Dispersion and Local Fuel Mixture Fraction}}.
\newblock \bibinfo{journal}{SAE International Journal of Engines}
  \DOIprefix\doi{10.4271/2011-01-0686}.
%Type = Article
\bibitem[{Post et~al.(1999)Post, Iyer and Abraham}]{Post2000}
\bibinfo{author}{Post, S.}, \bibinfo{author}{Iyer, V.},
  \bibinfo{author}{Abraham, J.}, \bibinfo{year}{1999}.
\newblock \bibinfo{title}{{A study of near-field entrainment in gas jets and
  sprays under diesel conditions}}.
\newblock \bibinfo{journal}{Journal of Fluids Engineering}
  \bibinfo{volume}{122}, \bibinfo{pages}{385--395}.
\newblock \DOIprefix\doi{10.1115/1.483268}.
%Type = Article
\bibitem[{Qiu and Reitz(2015)}]{Qiu2015}
\bibinfo{author}{Qiu, L.}, \bibinfo{author}{Reitz, R.D.}, \bibinfo{year}{2015}.
\newblock \bibinfo{title}{{An investigation of thermodynamic states during
  high-pressure fuel injection using equilibrium thermodynamics}}.
\newblock \bibinfo{journal}{International Journal of Multiphase Flow}
  \bibinfo{volume}{72}, \bibinfo{pages}{24--38}.
\newblock \DOIprefix\doi{10.1016/j.ijmultiphaseflow.2015.01.011}.
%Type = Article
\bibitem[{Ranz and Marshall(1952)}]{Ranz1952}
\bibinfo{author}{Ranz, W.E.}, \bibinfo{author}{Marshall, W.R.},
  \bibinfo{year}{1952}.
\newblock \bibinfo{title}{{Evaporation from drops. Parts I {\&} II.}}
\newblock \bibinfo{journal}{Chem. Eng. Progr} \bibinfo{volume}{48},
  \bibinfo{pages}{141--146; 173--180}.
\newblock \DOIprefix\doi{10.1016/S0924-7963(01)00032-X}.
%Type = Article
\bibitem[{Reitz(1987)}]{Reitz1987}
\bibinfo{author}{Reitz, R.D.}, \bibinfo{year}{1987}.
\newblock \bibinfo{title}{{Modeling Atomization Processes in High-Pressure
  Vaporizing Sprays}}.
\newblock \bibinfo{journal}{Atomization and Sprays} \bibinfo{volume}{3},
  \bibinfo{pages}{309--337}.
%Type = Article
\bibitem[{Reitz and Diwakar(1986)}]{Reitz1986}
\bibinfo{author}{Reitz, R.D.}, \bibinfo{author}{Diwakar, R.},
  \bibinfo{year}{1986}.
\newblock \bibinfo{title}{{Effect of Drop Breakup on Fuel Sprays}}.
\newblock \bibinfo{journal}{SAE Technical Paper 860469} ,
  \bibinfo{pages}{1--7}\DOIprefix\doi{10.4271/860469}.
%Type = Article
\bibitem[{Reitz and Diwakar(1987)}]{Reitz1987a}
\bibinfo{author}{Reitz, R.D.}, \bibinfo{author}{Diwakar, R.},
  \bibinfo{year}{1987}.
\newblock \bibinfo{title}{{Structure of High-Pressure Fuel Sprays}}.
\newblock \bibinfo{journal}{SAE Technical Paper 870598}
  \DOIprefix\doi{10.4271/870598}.
%Type = Manual
\bibitem[{Richards et~al.(2017)Richards, Senecal and Pomraning}]{Richards2017}
\bibinfo{author}{Richards, K.}, \bibinfo{author}{Senecal, P.},
  \bibinfo{author}{Pomraning, E.}, \bibinfo{year}{2017}.
\newblock \bibinfo{title}{{CONVERGE (v2.4)}}.
\newblock \bibinfo{organization}{Convergent Science Inc}.
  \bibinfo{address}{Madison, WI}.
%Type = Inproceedings
\bibitem[{Ritter et~al.(2017)Ritter, Korkmaz, Pitsch, Abel and
  Albin}]{Ritter2017}
\bibinfo{author}{Ritter, D.}, \bibinfo{author}{Korkmaz, M.},
  \bibinfo{author}{Pitsch, H.}, \bibinfo{author}{Abel, D.},
  \bibinfo{author}{Albin, T.}, \bibinfo{year}{2017}.
\newblock \bibinfo{title}{{Model-based Control of CNG-Diesel Dual-Fuel
  Engines}}, in: \bibinfo{booktitle}{AUTOREG 2017 – Automatisiertes Fahren
  und vernetzte Mobilität}, \bibinfo{publisher}{VDI/VDE Fachtagung},
  \bibinfo{address}{Berlin, Germany}.
%Type = Article
\bibitem[{Rosin and Rammler(1933)}]{Rosin1933}
\bibinfo{author}{Rosin, P.}, \bibinfo{author}{Rammler, E.},
  \bibinfo{year}{1933}.
\newblock \bibinfo{title}{{The Laws Governing the Fineness of powdered coal}}.
\newblock \bibinfo{journal}{Journal of the Institute of Fuel}
  \bibinfo{volume}{7}, \bibinfo{pages}{29--36}.
%Type = Article
\bibitem[{Rupp et~al.(2019)Rupp, Handschuh, Rieke and Kuperjans}]{Rupp2019}
\bibinfo{author}{Rupp, M.}, \bibinfo{author}{Handschuh, N.},
  \bibinfo{author}{Rieke, C.}, \bibinfo{author}{Kuperjans, I.},
  \bibinfo{year}{2019}.
\newblock \bibinfo{title}{{Contribution of country-specific electricity mix and
  charging time to environmental impact of battery electric vehicles: A case
  study of electric buses in Germany}}.
\newblock \bibinfo{journal}{Applied Energy} \bibinfo{volume}{237},
  \bibinfo{pages}{618--634}.
\newblock \DOIprefix\doi{10.1016/j.apenergy.2019.01.059}.
%Type = Article
\bibitem[{Rupp et~al.(2020)Rupp, Rieke, Handschuh and Kuperjans}]{Rupp2020}
\bibinfo{author}{Rupp, M.}, \bibinfo{author}{Rieke, C.},
  \bibinfo{author}{Handschuh, N.}, \bibinfo{author}{Kuperjans, I.},
  \bibinfo{year}{2020}.
\newblock \bibinfo{title}{{Economic and ecological optimization of electric bus
  charging considering variable electricity prices and CO2eq intensities}}.
\newblock \bibinfo{journal}{Transportation Research Part D: Transport and
  Environment} \bibinfo{volume}{81}.
\newblock \DOIprefix\doi{10.1016/j.trd.2020.102293}.
%Type = Article
\bibitem[{Rupp et~al.(2018)Rupp, Schulze and Kuperjans}]{Rupp2018}
\bibinfo{author}{Rupp, M.}, \bibinfo{author}{Schulze, S.},
  \bibinfo{author}{Kuperjans, I.}, \bibinfo{year}{2018}.
\newblock \bibinfo{title}{{Comparative Life Cycle Analysis of Conventional and
  Hybrid Heavy-Duty Trucks}}.
\newblock \bibinfo{journal}{World Electric Vehicle Journal}
  \bibinfo{volume}{9}, \bibinfo{pages}{33}.
\newblock \DOIprefix\doi{10.3390/wevj9020033}.
%Type = Article
\bibitem[{Rusanov(1961)}]{Rusanov1961}
\bibinfo{author}{Rusanov, V.V.}, \bibinfo{year}{1961}.
\newblock \bibinfo{title}{{Calculation of interaction of non--steady shock
  waves with obstacles}}.
\newblock \bibinfo{journal}{J. Comput. Math. Phys. USSR} .
%Type = Article
\bibitem[{Sazhin et~al.(2001)Sazhin, Feng and Heikal}]{Sazhin2001}
\bibinfo{author}{Sazhin, S.}, \bibinfo{author}{Feng, G.},
  \bibinfo{author}{Heikal, M.}, \bibinfo{year}{2001}.
\newblock \bibinfo{title}{{A model for fuel spray penetration}}.
\newblock \bibinfo{journal}{Fuel} \bibinfo{volume}{80},
  \bibinfo{pages}{2171--2180}.
\newblock \DOIprefix\doi{10.1016/S0016-2361(01)00098-9}.
%Type = Article
\bibitem[{Senecal and Leach(2019)}]{Senecal2019}
\bibinfo{author}{Senecal, P.K.}, \bibinfo{author}{Leach, F.},
  \bibinfo{year}{2019}.
\newblock \bibinfo{title}{{Diversity in transportation: Why a mix of propulsion
  technologies is the way forward for the future fleet}}.
\newblock \bibinfo{journal}{Results in Engineering} \bibinfo{volume}{4}.
\newblock \DOIprefix\doi{10.1016/j.rineng.2019.100060}.
%Type = Article
\bibitem[{Senecal et~al.(2014)Senecal, Pomraning, Richards and
  Som}]{Senecal2014}
\bibinfo{author}{Senecal, P.K.}, \bibinfo{author}{Pomraning, E.},
  \bibinfo{author}{Richards, K.J.}, \bibinfo{author}{Som, S.},
  \bibinfo{year}{2014}.
\newblock \bibinfo{title}{{Grid-Convergent Spray Models for Internal Combustion
  Engine Computational Fluid Dynamics Simulations}}.
\newblock \bibinfo{journal}{Journal of Energy Resources Technology}
  \bibinfo{volume}{136}, \bibinfo{pages}{012204}.
\newblock \DOIprefix\doi{10.1115/1.4024861}.
%Type = Article
\bibitem[{Siebers(1998)}]{Siebers1998}
\bibinfo{author}{Siebers, D.L.}, \bibinfo{year}{1998}.
\newblock \bibinfo{title}{{Liquid-phase fuel penetration in diesel sprays}}.
\newblock \bibinfo{journal}{SAE Technical Paper 980809}
  \DOIprefix\doi{10.4271/980809}.
%Type = Article
\bibitem[{Siebers(1999)}]{Siebers1999}
\bibinfo{author}{Siebers, D.L.}, \bibinfo{year}{1999}.
\newblock \bibinfo{title}{{Scaling liquid-phase fuel penetration in diesel
  sprays based on mixing-limited vaporization}}.
\newblock \bibinfo{journal}{SAE Technical Paper 1999-01-0528}
  \DOIprefix\doi{10.4271/1999-01-0528}.
%Type = Article
\bibitem[{Skeen et~al.(2015)Skeen, Manin and Pickett}]{Skeen2015}
\bibinfo{author}{Skeen, S.A.}, \bibinfo{author}{Manin, J.},
  \bibinfo{author}{Pickett, L.M.}, \bibinfo{year}{2015}.
\newblock \bibinfo{title}{{Simultaneous formaldehyde PLIF and high-speed
  schlieren imaging for ignition visualization in high-pressure spray flames}}.
\newblock \bibinfo{journal}{Proceedings of the Combustion Institute}
  \bibinfo{volume}{35}, \bibinfo{pages}{3167--3174}.
\newblock \DOIprefix\doi{10.1016/j.proci.2014.06.040}.
%Type = Article
\bibitem[{Som et~al.(2011)Som, Ramirez, Longman and Aggarwal}]{Som2011}
\bibinfo{author}{Som, S.}, \bibinfo{author}{Ramirez, A.I.},
  \bibinfo{author}{Longman, D.E.}, \bibinfo{author}{Aggarwal, S.K.},
  \bibinfo{year}{2011}.
\newblock \bibinfo{title}{{Effect of nozzle orifice geometry on spray,
  combustion, and emission characteristics under diesel engine conditions}}.
\newblock \bibinfo{journal}{Fuel} \bibinfo{volume}{90},
  \bibinfo{pages}{1267--1276}.
\newblock \DOIprefix\doi{10.1016/j.fuel.2010.10.048}.
%Type = Book
\bibitem[{{U. S. Energy Information Administration}(2016)}]{USEIA2016}
\bibinfo{author}{{U. S. Energy Information Administration}},
  \bibinfo{year}{2016}.
\newblock \bibinfo{title}{{International Energy Outlook 2016}}.
%Type = Article
\bibitem[{{von Kuensberg Sarre} et~al.(1999){von Kuensberg Sarre}, Kong and
  Reitz}]{Sarre1999}
\bibinfo{author}{{von Kuensberg Sarre}, C.}, \bibinfo{author}{Kong, S.c.},
  \bibinfo{author}{Reitz, R.D.}, \bibinfo{year}{1999}.
\newblock \bibinfo{title}{{Modeling the Effects of Injector Nozzle Geometry on
  Diesel Sprays}}.
\newblock \bibinfo{journal}{SAE Technical Paper 1999-01-0912} ,
  \bibinfo{pages}{1--14}\DOIprefix\doi{10.4271/1999-01-0912}.
%Type = Article
\bibitem[{Wakuri et~al.(1959)Wakuri, Fujii, Amitani and Tsuneya}]{Wakuri1959}
\bibinfo{author}{Wakuri, Y.}, \bibinfo{author}{Fujii, M.},
  \bibinfo{author}{Amitani, T.}, \bibinfo{author}{Tsuneya, R.},
  \bibinfo{year}{1959}.
\newblock \bibinfo{title}{{Studies on the Penetration of Fuel Spray of Diesel
  Engine}}.
\newblock \bibinfo{journal}{Transactions of the Japan Society of Mechanical
  Engineers} \bibinfo{volume}{25}, \bibinfo{pages}{820--826}.
\newblock \DOIprefix\doi{10.1299/kikai1938.25.820}.
%Type = Article
\bibitem[{Wakuri et~al.(1960)Wakuri, Fujii, Amitani and Tsuneya}]{Wakuri1960}
\bibinfo{author}{Wakuri, Y.}, \bibinfo{author}{Fujii, M.},
  \bibinfo{author}{Amitani, T.}, \bibinfo{author}{Tsuneya, R.},
  \bibinfo{year}{1960}.
\newblock \bibinfo{title}{{Studies on the Penetration of Fuel Spray in a Diesel
  Engine}}.
\newblock \bibinfo{journal}{Bulletin of JSME} \bibinfo{volume}{3},
  \bibinfo{pages}{123--130}.
\newblock \DOIprefix\doi{10.1299/jsme1958.3.123}.
%Type = Book
\bibitem[{Wallis(1969)}]{Wallis1969}
\bibinfo{author}{Wallis, G.B.}, \bibinfo{year}{1969}.
\newblock \bibinfo{title}{{One-Dimensional Two-Phase Flow}}.
\newblock \bibinfo{publisher}{McGraw-Hill, New York}.
%Type = Phdthesis
\bibitem[{Wan(1997)}]{Wan1997}
\bibinfo{author}{Wan, Y.}, \bibinfo{year}{1997}.
\newblock \bibinfo{title}{{Numerical Study of Transient Fuel Sprays with
  Autoignition and Combustion Under Diesel-Engine Relevant Conditions}}.
\newblock Ph.D. thesis. RWTH Aachen University.
%Type = Article
\bibitem[{Wan and Peters(1999)}]{Wan1999}
\bibinfo{author}{Wan, Y.}, \bibinfo{author}{Peters, N.}, \bibinfo{year}{1999}.
\newblock \bibinfo{title}{{Scaling of Spray Penetration with Evaporation}}.
\newblock \bibinfo{journal}{Atomization and Sprays} \bibinfo{volume}{9},
  \bibinfo{pages}{111--132}.
\newblock \DOIprefix\doi{10.1615/AtomizSpr.v9.i2.10}.
%Type = Article
\bibitem[{Wan and Peters(1997)}]{Wan1997a}
\bibinfo{author}{Wan, Y.P.}, \bibinfo{author}{Peters, N.},
  \bibinfo{year}{1997}.
\newblock \bibinfo{title}{{Application of the Cross-Sectional Average Method to
  Calculations of the Dense Spray Region in a Diesel Engine}}.
\newblock \bibinfo{journal}{SAE Technical Paper 972866}
  \DOIprefix\doi{10.4271/972866}.
%Type = Inproceedings
\bibitem[{Wang et~al.(2014)Wang, Raju, Pomraning, Kundu, Pei and
  Som}]{Wang2014}
\bibinfo{author}{Wang, M.}, \bibinfo{author}{Raju, M.},
  \bibinfo{author}{Pomraning, E.}, \bibinfo{author}{Kundu, P.},
  \bibinfo{author}{Pei, Y.}, \bibinfo{author}{Som, S.}, \bibinfo{year}{2014}.
\newblock \bibinfo{title}{{Comparison of Representative Interactive Flamelet
  and Detailed Chemistry Based Combustion Models for Internal Combustion
  Engines}}, in: \bibinfo{booktitle}{Volume 2: Instrumentation, Controls, and
  Hybrids; Numerical Simulation; Engine Design and Mechanical Development;
  Keynote Papers}, \bibinfo{publisher}{American Society of Mechanical
  Engineers}.
\newblock \DOIprefix\doi{10.1115/ICEF2014-5522}.
%Type = Inproceedings
\bibitem[{Wehrfritz et~al.(2012)Wehrfritz, Vuorinen, Kaario and
  Larmi}]{Wehrfritz2012}
\bibinfo{author}{Wehrfritz, A.}, \bibinfo{author}{Vuorinen, V.},
  \bibinfo{author}{Kaario, O.}, \bibinfo{author}{Larmi, M.},
  \bibinfo{year}{2012}.
\newblock \bibinfo{title}{{A High Resolution Study of Non-Reacting Fuel Sprays
  Using Large-Eddy Simulations}}, in: \bibinfo{booktitle}{ICLASS 2012},
  \bibinfo{address}{Heidelberg, Germany}.
%Type = Article
\bibitem[{Wei et~al.(2017)Wei, Chen, Zhao, Zhou and Chen}]{Wei2017}
\bibinfo{author}{Wei, H.}, \bibinfo{author}{Chen, X.}, \bibinfo{author}{Zhao,
  W.}, \bibinfo{author}{Zhou, L.}, \bibinfo{author}{Chen, R.},
  \bibinfo{year}{2017}.
\newblock \bibinfo{title}{{Effects of the turbulence model and the spray model
  on predictions of the n-heptane jet fuel–air mixing and the ignition
  characteristics with a reduced chemistry mechanism}}.
\newblock \bibinfo{journal}{Proceedings of the Institution of Mechanical
  Engineers, Part D: Journal of Automobile Engineering} \bibinfo{volume}{231},
  \bibinfo{pages}{1877--1888}.
\newblock \DOIprefix\doi{10.1177/0954407016687478}.
%Type = Article
\bibitem[{Wilke(1950)}]{Wilke1950}
\bibinfo{author}{Wilke, C.R.}, \bibinfo{year}{1950}.
\newblock \bibinfo{title}{{A viscosity equation for gas mixtures}}.
\newblock \bibinfo{journal}{The Journal of Chemical Physics}
  \bibinfo{volume}{18}, \bibinfo{pages}{517}.
\newblock \DOIprefix\doi{10.1063/1.1747673}.
%Type = Article
\bibitem[{Xu et~al.(2016)Xu, Cui and Deng}]{Xu2016}
\bibinfo{author}{Xu, M.}, \bibinfo{author}{Cui, Y.}, \bibinfo{author}{Deng,
  K.}, \bibinfo{year}{2016}.
\newblock \bibinfo{title}{{One-dimensional model on liquid-phase fuel
  penetration in diesel sprays}}.
\newblock \bibinfo{journal}{Journal of the Energy Institute}
  \bibinfo{volume}{89}, \bibinfo{pages}{138--149}.
\newblock \DOIprefix\doi{10.1016/j.joei.2015.01.002}.
%Type = Article
\bibitem[{Xue et~al.(2013)Xue, Som, Senecal and Pomraning}]{Xue2013}
\bibinfo{author}{Xue, Q.}, \bibinfo{author}{Som, S.}, \bibinfo{author}{Senecal,
  P.K.}, \bibinfo{author}{Pomraning, E.}, \bibinfo{year}{2013}.
\newblock \bibinfo{title}{{Large eddy simulation of fuel-spray under
  non-reacting ic engine conditions}}.
\newblock \bibinfo{journal}{Atomization and Sprays} \bibinfo{volume}{23},
  \bibinfo{pages}{925--955}.
\newblock \DOIprefix\doi{10.1615/AtomizSpr.2013008320}.
%Type = Article
\bibitem[{Zhou et~al.(2011)Zhou, Xie, Jia and Shi}]{Zhou2011}
\bibinfo{author}{Zhou, L.}, \bibinfo{author}{Xie, M.Z.}, \bibinfo{author}{Jia,
  M.}, \bibinfo{author}{Shi, J.R.}, \bibinfo{year}{2011}.
\newblock \bibinfo{title}{{Large eddy simulation of fuel injection and mixing
  process in a diesel engine}}.
\newblock \bibinfo{journal}{Acta Mechanica Sinica} \bibinfo{volume}{27},
  \bibinfo{pages}{519}.
\newblock \DOIprefix\doi{10.1007/s10409-011-0485-1}.

\end{thebibliography}

%\printacronyms[include-classes=abbrev,name=Abbreviations]

%\printacronyms[include-classes=nomencl,name=Nomenclature]

%\printacronyms[include-classes=subscr,name=Subscripts]

%\printacronyms[include-classes=supscr,name=Superscripts]

%\Appendix if required

\end{document}